\documentclass[aps,amsmath,amssymb, notitlepage, 
footinbib,
twocolumn, superscriptaddress
]{revtex4-1}

\usepackage{graphicx}
\usepackage{color}
\usepackage{bbold}

\newcommand{\bit}{\begin{itemize}}
\newcommand{\eit}{\end{itemize}}

\newcommand{\f}{\frac}
\renewcommand{\>}{\right\rangle}
\newcommand{\<}{\left\langle}
\newcommand{\ba}{\begin{align}}
\newcommand{\ea}{\end{align}}
\newcommand{\be}{\begin{equation}}
\newcommand{\ee}{\end{equation}}
\newcommand{\bi}{\begin{itemize}}
\newcommand{\ei}{\end{itemize}}
\newcommand{\lf}{\left(}
\newcommand{\ri}{\right)}
\newcommand{\dd}{\mathrm{d}}

\newcommand{\Tr}{\operatorname{Tr}}
\newcommand{\tr}{\operatorname{tr}}

\newcommand{\rp}{\mathbb{RP}}
\newcommand{\cp}{\mathbb{CP}}
\newcommand{\kt}{\widetilde K}
\newcommand{\q}{\mathrm{q}}

\begin{document}

\newcommand{\bra}[1]{\< #1 \right|}
\newcommand{\ket}[1]{\left| #1 \>}

\title{Loop models with crossings}

\author{Adam Nahum}
\affiliation{Theoretical Physics, Oxford University, 1 Keble Road, Oxford OX1 3NP, United Kingdom}
\author{P. Serna} 
\affiliation{Departamento de F\'isica -- CIOyN, Universidad de Murcia, Murcia 30.071, Spain}
\author{A. M. Somoza}
\affiliation{Departamento de F\'isica -- CIOyN, Universidad de Murcia, Murcia 30.071, Spain}
\author{M. Ortu\~no}
\affiliation{Departamento de F\'isica -- CIOyN, Universidad de Murcia, Murcia 30.071, Spain}
\date{\today}

\begin{abstract}
The universal behaviour of two-dimensional loop models can change dramatically when loops are allowed to cross. We study models with crossings both analytically and with extensive Monte Carlo simulations. Our main focus (the `completely packed loop model with crossings') is a simple generalisation of well-known models  which shows an interesting phase diagram with continuous phase transitions of a new kind. These separate the unusual `Goldstone' phase observed previously from phases with short loops. Using mappings to $\mathbb{Z}_2$ lattice gauge theory, we show that the continuum description of the model is a replica limit of the sigma model on real projective space ($\mathbb{RP}^{n-1}$). This field theory sustains $\mathbb{Z}_2$ point defects which proliferate at the transition.  In addition to studying the new critical points, we characterise the universal properties of the Goldstone phase in detail, comparing renormalisation group (RG) calculations with numerical data on systems of linear size up to $L=10^6$ at loop fugacity $n=1$. (Very large  sizes are necessary because of the logarithmic form of correlation functions and other observables.)  The model is relevant to polymers on the verge of collapse, and a particular point in parameter space maps to  self-avoiding trails at their $\Theta$-point; we use the RG treatment of a perturbed sigma model to resolve some perplexing features in the previous literature on trails.  Finally, one of the phase transitions considered here is a close analogue of those in disordered electronic systems  --- specifically, Anderson metal-insulator transitions --- and  provides a simpler context in which to study the properties of these poorly-understood (central-charge-zero) critical points.\\
\end{abstract}

\maketitle

\section{Introduction}

\begin{figure}[t] 
\centering
\includegraphics[width=3.44in]{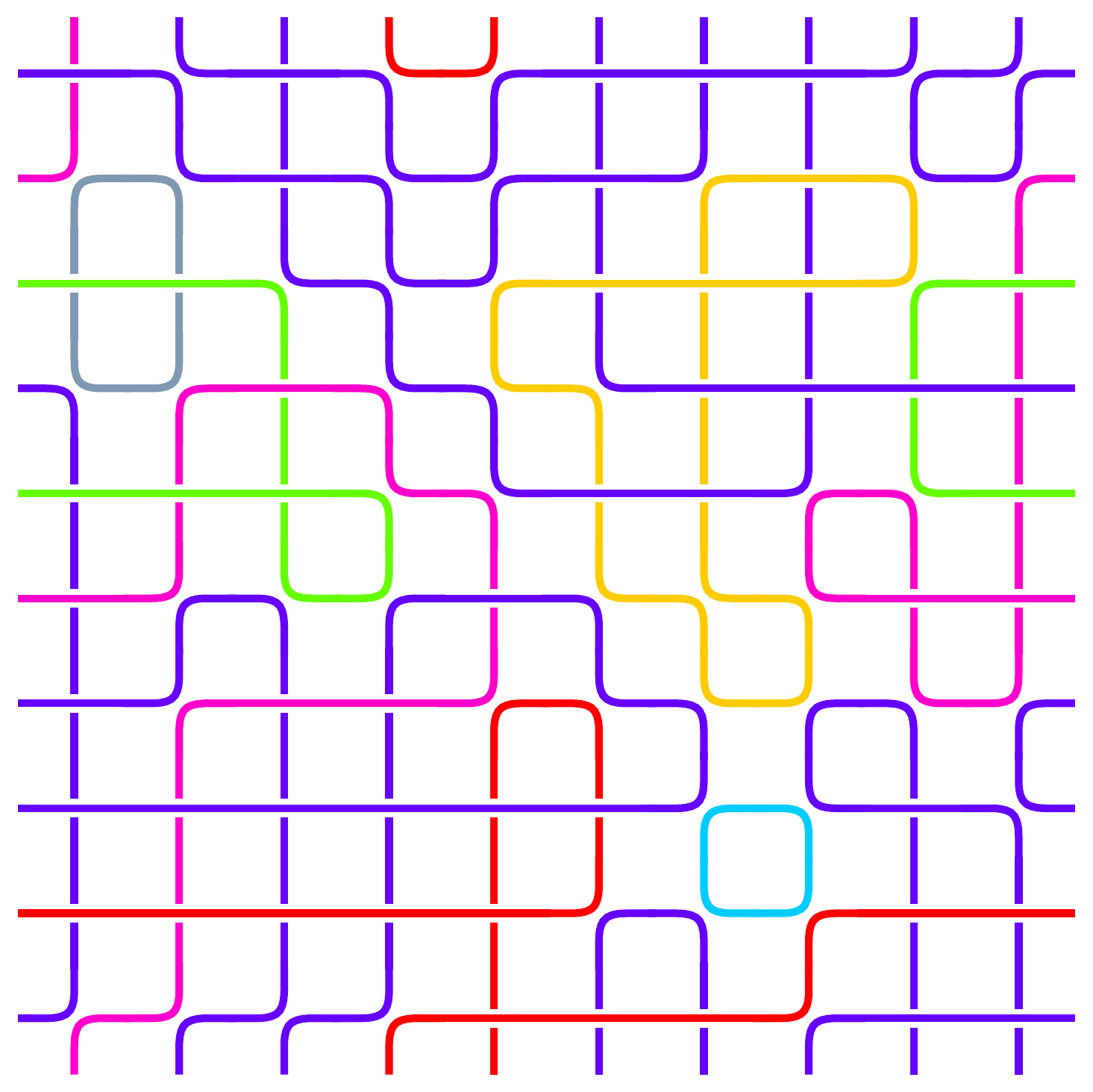}
\caption{A configuration of the completely packed loop model with crossings (CPLC) on a $10 \times 10$ lattice with periodic boundary conditions. Each loop has been given a different colour -- these colours are merely a guide to the eye and not part of the configuration.}
\label{7looppicture}
\end{figure}

It is tempting to think that two-dimensional critical phenomena are completely classified and understood, thanks to conformal field theory and other exact techniques, but this is far from true. One class of problems which remains mysterious is that containing classical loop models and models for polymers, together with models for noninteracting fermions subject to disorder. These systems are tied together by field theory descriptions with continuous replica-like symmetries (or alternatively global supersymmetries). Examples include de Gennes' mapping of polymers to the $O(N)$ model in the limit $N\rightarrow 0$, the various replica sigma models for the integer quantum Hall transition and other localisation problems, and the sigma models describing cluster boundaries in percolation and similar soups of loops \cite{replica for loops,Pruisken, Levine et al,McKane, Parisi Sourlas SUSY for loops, Weidenmuller, mirlin evers review, Fendley sigma models review, candu et al, read saleur, dense loops and supersymmetry, vortex paper, short loop paper}.
 
Classical loop models yield the simplest examples of this class of problems, but even they are not yet fully understood. The best-studied examples are those in which the loops are forbidden from crossing: for these a great deal is known from conformal field theory,  height model mappings, exact solutions,  Schramm-Loewner Evolution, and numerical simulations \cite{Nienhuis review, Cardy SLE review}. But when we move away from these models the analytical techniques often cease to apply, and we may encounter new types of critical phenomena requiring new theoretical tools. 

This paper considers some of the simplest two-dimensional loop models with crossings. These  reveal new universality classes of, and new mechanisms for, classical critical behaviour. They also provide natural models for polymers  and for deterministic motion in a random environment \cite{Owczarek and Prellberg collapse, foster universality, Ziff} which have been intensely studied but whose phase diagrams and continuum descriptions have in general not been understood. Finally, they shed light on phenomena that are important more generally for criticality in replica or supersymmetric sigma models -- in particular, the role of gauge symmetries and topological point defects. The latter have recently been shown also to be important for two-dimensional Anderson metal-insulator transitions \cite{Konig et al, Fu Kane}. We will return to the analogy between loop models and localisation at the end of this introduction.

A key result of previous work on loops with crossings is the existence of an unusual critical phase which is absent for non-crossing loops \cite{dense loops and supersymmetry, read saleur, Ziff, Owczarek and Prellberg collapse, Martins et al, Kager Nienhuis, Ikhlef nonintersection}. It was argued by Jacobsen, Read and Saleur \cite{read saleur, dense loops and supersymmetry} that this corresponds to the \emph{Goldstone} phase of the $O(n)$ sigma model, where $n$ is the fugacity for loops. The phase exists for $n<2$; to make sense of this regime requires a replica-like limit or a supersymmetric formulation of the field theory. Characteristic features of the Goldstone phase had previously been found in computational studies of polymers and deterministic walks in a random environment \cite{Ziff, Owczarek and Prellberg collapse}, as well as in an integrable loop model \cite{Martins et al, Kager Nienhuis, Ikhlef nonintersection}. The phase appears quite generically when non-crossing loop models in the so-called `dense' regime are perturbed by the addition of crossings, which in an appropriate field theory corresponds to a breaking of symmetry \cite{dense loops and supersymmetry, read saleur, vortex paper}.

Here we examine a more general class of loop models with crossings. These show new continuous phase transitions separating the Goldstone phase from non-critical phases with short loops.  We construct field theories for the models, and pin down the universal behaviour (both in the Goldstone phase and at the new critical points) using analytic calculations and extensive Monte Carlo simulations. Finally, we give a field theoretic treatment of the closely related problem of the interacting self-avoiding trail model for a polymer.

The models we study are a `completely-packed loop model with crossings' (CPLC) on the square lattice ---  Fig.~\ref{7looppicture} shows a configuration --- and an `incompletely packed loop model with crossings' (IPLC)  in which loops are related to cluster boundaries. The parameter space for the CPLC contains various previously-studied models as special cases: in particular, the standard completely-packed loop model \emph{without} crossings and models with crossings encountered in various contexts (including those mentioned above). The CPLC and IPLC are expected to show the same universal behaviour --- our numerics are restricted to the CPLC, but the IPLC provides a simpler context in which to describe the main theoretical ideas.

It is easy to argue that the phase transitions in the CPLC and IPLC cannot be described by the $O(n)$ model. Instead, the general description of the models requires couping the $O(n)$ spin $\vec S$ to a $\mathbb{Z}_2$ gauge field, or equivalently identifying $\vec S$ with  $-\vec S$ to obtain a nematic order parameter. This leads to a sigma model on real projective space, $\rp^{n-1}$, in which $\mathbb{Z}_2$ point vortices play an important role. Vortices are suppressed in the Goldstone phase --- meaning that the $O(n)$ model is a viable description there --- but proliferate at the phase transition into the short loop phase. (This picture is appropriate for the regime $0<n<2$.)

In general, the introduction of crossings leads the standard exact techniques used for non-crossing loop models to fail, so for the critical points we are restricted to numerics and approximate RG treatments \cite{Fu Kane}. However the Goldstone phase can be fully understood analytically, since it is characterised by marginal flow to a weak-coupling fixed point \cite{dense loops and supersymmetry, read saleur}. This leads to logarithms --- e.g. correlation functions decaying with a universal power of the logarithm of distance --- so very large system sizes are required in order to confirm our analytical predictions numerically  (comparable to the largest sizes simulated in any statistical mechanics problem). These are possible at fugacity $n=1$ thanks to special features of the problem there, and our simulations are restricted to this value.

Another feature of the CPLC at $n=1$ is that while each configuration is a soup of many loops, the model permits a mapping to a model for a \emph{single} loop with local interactions. At a certain point in parameter space, this is the well-studied `interacting self-avoiding trail' (ISAT) model for a polymer at its collapse, or $\Theta$, point \cite{Lyklema, Owczarek and Prellberg collapse}. Collapse transitions for polymers in two dimensions are a mysterious subject  which deserves clarification (see for instance the case of the missing Flory exponents \cite{exact scaling functions}). Here we show that the ISAT can be understood completely from field theory, explaining for example the interesting phase diagram found numerically in Ref.~\cite{foster universality}. To do this we perturb the sigma model that describes the Goldstone phase. Surprisingly, the $\Theta$ point of the ISAT turns out to be an \emph{infinite order} multicritical point: despite the simplicity and naturalness of this model, it is highly fine-tuned from the point of view of general polymer models. This implies that the critical exponents for the  \emph{generic} $\Theta$ point polymer (with crossings) are still unknown.

There is a close relationship between loop models at loop fugacity $n=1$ and disordered fermion problems \cite{glr, beamond cardy chalker, ortuno somoza chalker, conformal restriction, modified CC model, Mirlin Evers Mildenberger, cardy network models review}. Supersymmetry and replica-like limits, crucial in the latter for averaging over disorder, appear in the former as tools allowing geometrical correlation functions (such as the probability that two points lie on the same loop) to be expressed in field theory. Both types of problem exhibit critical points of central charge zero, described by logarithmic CFTs \cite{gurarie logs, log cft reviews}. The loop models are a good place to study such critical points since they are more tractable, both analytically and computationally, than disordered fermion problems.

For completely-packed loops without crossings, there is in fact an exact mapping \cite{glr, beamond cardy chalker} to a network model for Anderson localization in symmetry class C \cite{quasiparticle localisation in high Tc, SQHE in unconventional superconductors, mirlin evers review}. However the analogy is more general. Recent work by Fu and Kane \cite{Fu Kane} demonstrates that the metal-insulator transition in the symplectic symmetry class is driven by proliferation of $\mathbb{Z}_2$ vortices: this transition is thus in remarkably close analogy with those in the loop models discussed here, though the appropriate sigma model is different. In the localisation language, the Goldstone phase corresponds to a metallic phase, and the two short loop phases --- which are distinguished from each other by the presence or absence of a loop encircling the boundary ---  to topological and trivial insulating phases.

The CPLC at $n=1$ can in fact be obtained as a `classical' limit of a network model in which a Kramers doublet propagates on every edge: the above similarities show that this classical limit captures a surprising number of the qualitative features of the phase diagram for the symplectic class.

In both the loop model and the localisation problem the vortex fugacity (including its sign) plays an important role. Ref.~\cite{Fu Kane} introduced an approximate RG treatment of this fugacity, and in Sec.~\ref{RG equations with vortices} we apply this to the loop models. We note that vortices --- this time $\mathbb{Z}$ vortices --- have also been shown to be responsible for Anderson localisation in the chiral symmetry classes, and a detailed treatment has been given by K\"onig et al. in Ref.~\cite{Konig et al}.

The organisation of the paper is as follows. In the next section we introduce the models we will study and their phase diagrams. In Sec.~\ref{lattice field theory section} we map them to lattice gauge theories and $\rp^{n-1}$ sigma models, paying attention to the role of topological defects and the relation to the field theory for loops without crossings. We also discuss the different ($\cp^{n-1}$) sigma model which applies on the boundaries of the phase diagram for the CPLC. Sections \ref{Goldstone phase} and \ref{critical line} apply Monte Carlo and RG calculations in the field theory to the Goldstone phase and the critical points respectively. Our numerical methods are described in more detail in Sec.~\ref{Numerical methods}. Sec.~\ref{polymer section} tackles the polymer collapse problem. Finally, Sec.~\ref{conclusions} discusses directions for future work.

\section{Definitions of models}

\subsection{Completely-packed loops with crossings}
\label{CPLC}

\begin{figure}[b] 
\centering
\includegraphics[width=3.2in]{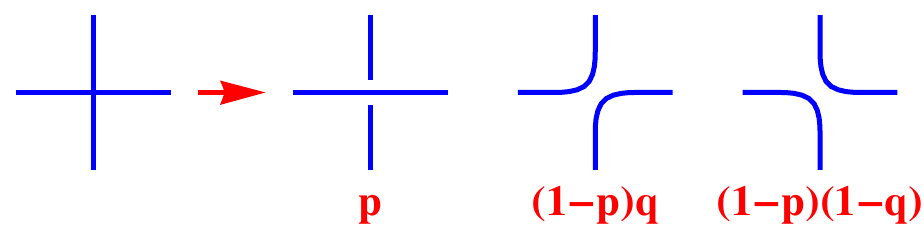}
\caption{The three configurations of a node and associated Boltzmann weights. (In the leftmost configuration, the upper and lower links lie on the same loop.) The weights $q$ and $1-q$ are exchanged on the two sublattices of the square lattice.}
\label{decompnode}
\end{figure}

Configurations of the completely packed loop model with crossings (CPLC) are generated by resolving each node of the square lattice in one of the three possible ways shown in Fig.~\ref{decompnode}. Fig.~\ref{7looppicture} shows an example on a small lattice. Note that  overcrossings are not distinguished from undercrossings --- the configuration at a node is defined solely by the way its four links are paired up. 

\begin{figure}[t] 
\centering
\includegraphics[width=3.1in]{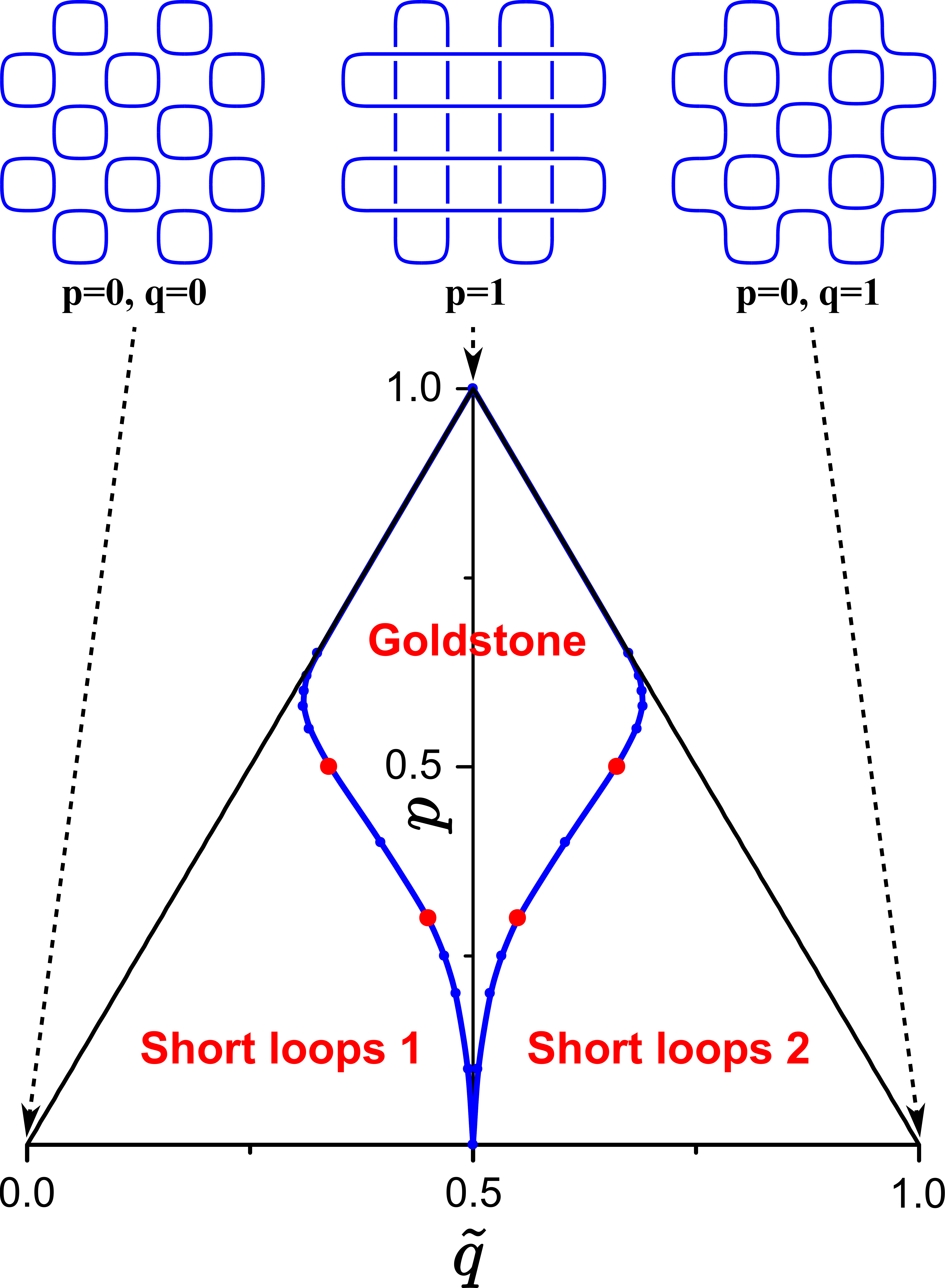}
\caption{Phase diagram obtained numerically for the CPLC at $n=1$. The horizontal axis is labelled by $\widetilde q$, defined by $(\widetilde q-1/2) = (q-1/2)(1-p)$. The larger (red) dots on the critical line indicate the values of $p$ at which we have analysed the critical behaviour in detail. Also shown are the configurations obtaining on a small finite lattice at $p=1$, at $p=0$, $q=0$, and at $p=0$, $q=1$. (The point $p=0$, $q=1/2$ is the percolation critical point.)}
\label{phasediagram}
\end{figure}

Each of the three possible pairings at a node is assigned a weight, as shown in Fig.~\ref{decompnode}, with the weight of a crossing being $p$. The factors $q$ and $1-q$ are staggered (swapped) on the two sublattices of the square lattice, so that the states of the system for extreme values of the parameters are as shown in Fig.~\ref{phasediagram}. The Boltzmann weight for a configuration is given by the product of the node factors, together with a fugacity $n$ for the loops. Let $N_p$, $N_q$ and $N_{1-q}$ denote the numbers of nodes where the pairing with weight $p$, $(1-p)q$ or $(1-p)(1-q)$ is chosen. Then the product of node weights in a configuration $\mathcal{C}$ is
 \ba \label{W weight}
W_\mathcal{C}  =  p^{N_p} \left[ (1-p) q \right]^{N_q} \left[ (1-p) (1-q) \right]^{N_{1-q}},
\end{align}
and the partition function is
\ba \label{ZCPLC}
Z  =
\sum_{\mathcal{C}}
    \, n^\text{no. loops}  \, W_\mathcal{C}.
 \end{align}

The parameter space of this model includes various previously-investigated models. On the line $p=0$ we have the completely-packed loop model \emph{without} crossings, which has been intensely studied and which may be mapped to the $n^2$-state Potts model via the Fortuin-Kasteleyn representation of the latter.  The model on the line $q=1/2$ was related to the Goldstone phase of the $O(n)$ sigma model in Ref.~\cite{dense loops and supersymmetry}, and points on this line have been studied in various contexts. For a given value of $n$ the point $q=1/2$, $p=(2-n)/(10-n)$ is known as the Brauer loop model \cite{Martins et al, Kager Nienhuis, Ikhlef nonintersection} and is integrable;  this model was related to a supersymmetric spin chain in Ref.~\cite{Martins et al}. When the parameters in the CPLC are such that all configurations are given equal weight --- i.e. when $n=1$, $q=1/2$ and $p=1/3$ --- it is equivalent to a standard model for polymers at their $\Theta$  (collapse) point \cite{Lyklema, Owczarek and Prellberg collapse, foster universality}, which we will discuss further in  Sec.~\ref{polymer section}. On the lines $q=0$ or $q=1$ --- the left and right boundaries of the phase diagram Fig.~\ref{phasediagram} --- the  CPLC reduces to the `Manhattan' lattice loop model discussed in \cite{beamond cardy chalker, beamond owczarek cardy}.  Loop models with crossings at $n=1$ have also appeared in the study of Lorentz lattice gases, i.e. deterministic motion in a random environment \cite{Ziff, Gunn Ortuno}. Finally, Ref.~\cite{Shtengel Chayes} discusses a model similar to the CPLC in which $q$ and $1-q$ are not staggered, and uses it (at $n=2$) to analyse the phase diagrams of vertex models.

A trivial but important fact about the CPLC is that the nodes become completely independent of each other when $n=1$. The weights $p$, $(1-p)q$ and $(1-p)(1-q)$ are then the probabilities of the various node configurations,  and the partition function $Z$ is equal to unity --- from which it follows, by the finite size scaling of the free energy, that any critical points  must have central charge $c=0$.  The model with $n=1$ is thus analogous to percolation, which can also be formulated in terms of uncorrelated random variables. In the absence of crossings the $n=1$  model is in fact equivalent to bond percolation on a dual lattice, with loops surrounding cluster boundaries; however when crossings are allowed the universal behaviour is no longer that of percolation.

Our simulations will be restricted to the case $n=1$, which is  the most interesting and the best suited to Monte Carlo, but most analytic results will apply to $0\leq n <2$. (We will discuss the case $n=2$, which shows more conventional critical behaviour, elsewhere.) The phase diagram obtained numerically at $n=1$ is shown in Fig.~\ref{phasediagram}. We expect it to be qualitatively similar for $0<n<2$, with the Goldstone phase swallowing up more and more of the parameter space as $n\rightarrow 0$. We now summarize its main features.

{\bf Short loop phases.} The configurations at $p=0$, $q=0$ and at $p=0$, $q=1$ provide caricatures of the two `short loop' phases. For given boundary conditions, these are distinguished from each other by the presence or absence of a long loop running along the boundary, as shown in Fig.~\ref{phasediagram}. In a sigma model description the short loop phases are {massive} (disordered) phases. In the analogy with Anderson localization mentioned in the introduction, they correspond to insulating phases, and the boundary loops correspond to the edge states present in a topological insulator.

{\bf Goldstone phase.} In the Goldstone phase, so-called because the continuum description is a  sigma model which flows to weak coupling in the infra-red \cite{dense loops and supersymmetry}, the loops are `almost' Brownian. However, interactions between  Goldstone modes in the sigma model are only marginally irrelevant, leading to universal logarithmic forms for correlators and other observables which we calculate in Sec.~\ref{Goldstone phase}. In the Anderson localization analogy, this would be a metallic phase.

{\bf Critical lines.} The lines separating the Goldstone phase from the short-loop phases show a new universality class of critical behaviour. This is associated with the order-disorder transition of the $\rp^{n-1}$ sigma model, which exists only in the replica limit $n<2$ and is driven by proliferation of $\mathbb{Z}_2$ vortex defects associated with $\pi_1(\mathbb{RP}^{n-1})$ (Secs.~\ref{lattice field theory section},~\ref{RG equations with vortices}). Numerically, the critical loops have $d_f = 1.909(1)$ at $n=1$, i.e. they are slightly less compact than Brownian paths, and the transition has a large correlation length exponent $\nu = 2.745(19)$ (Sec.~\ref{critical line}).

{\bf Critical point at $p=0$.} The critical point of the loop model without crossings (at $p=0$, $q=1/2$) is well-studied and corresponds to the so-called dense phase of the $O(n)$ loop model, or to SLE$\phantom{}_\kappa$ with $\kappa > 4$. (Note that the standard terminology `$O(n)$ loop model' is potentially misleading, as when crossings are forbidden the $O(n)$ model is not the appropriate field theory for the dense phase \cite{dense loops and supersymmetry, candu et al}.) At $n=1$ this point maps to critical percolation:  the loops have the statistics of  percolation cluster boundaries, with a fractal dimension $d_f^\text{perc} = 7/4$, and the correlation length exponent of the transition is $\nu^\text{perc} = 4/3$. These critical exponents  also yield exponents in an Anderson transition (the spin quantum Hall transition)  via an exact mapping \cite{glr, beamond cardy chalker, Mirlin Evers Mildenberger, chalker ortuno somoza}.

{\bf Phase diagram boundaries.} Everywhere on the boundary of the phase diagram Fig.~\ref{phasediagram} --- i.e. whenever one of the node weights vanishes --- loops can be {consistently oriented} by assigning a \emph{fixed} (configuration independent) orientation to each link of the lattice. The necessary choice of  orientations differs on each of the three pieces of boundary. They are those of the `L' lattice on the line $p=0$, and of the Manhattan lattice on the lines $q=0$ and $q=1$. (These lattices are depicted in Ref.~\cite{beamond cardy chalker}.) The fact that the loops  automatically come with an orientation means that the continuum descriptions have a higher symmetry \cite{read saleur enlarged symm, candu et al}, as will be discussed in Sec.~\ref{boundaries of phase diagram}, and the appropriate field theory is a sigma model on complex projective space, $\mathbb{CP}^{n-1}$, rather than on $\mathbb{RP}^{n-1}$. The $\mathbb{CP}^{n-1}$ description  implies that the lines $q=0$ and $q=1$ (the Manhattan lattice loop model) are always in the short loop phase, but with a typical loop length that diverges exponentially as $p\rightarrow 1$ (Sec.~\ref{boundaries of phase diagram}). This is in agreement with previous expectations  \cite{beamond owczarek cardy, beamond cardy chalker}, but is not obvious from the numerical phase diagram since the critical lines closely approach the lines $q=0,1$ for $p$ close to one.

{\bf Relation to polymers.} Configurations in the CPLC are soups of many loops. However when $n=1$ the CPLC has a simple relation with the self-avoiding trail model for a single polymer \cite{Owczarek and Prellberg collapse, foster universality}. The polymer corresponds to a single marked loop in the soup of loops; so long as $n=1$, `integrating out' the configurations of the other loops leads to a local Boltzmann weight for the marked one. Adding the interactions that are natural in the polymer language takes us out of the parameter space of Fig.~\ref{phasediagram}, but the sigma model description can be extended to cover this case by including appropriate symmetry-breaking terms (Sec.~\ref{polymer section}).

\subsection{Incompletely-packed loops with crossings}
\label{second model}

\begin{figure}[t] 
\centering
\includegraphics[width=3in]{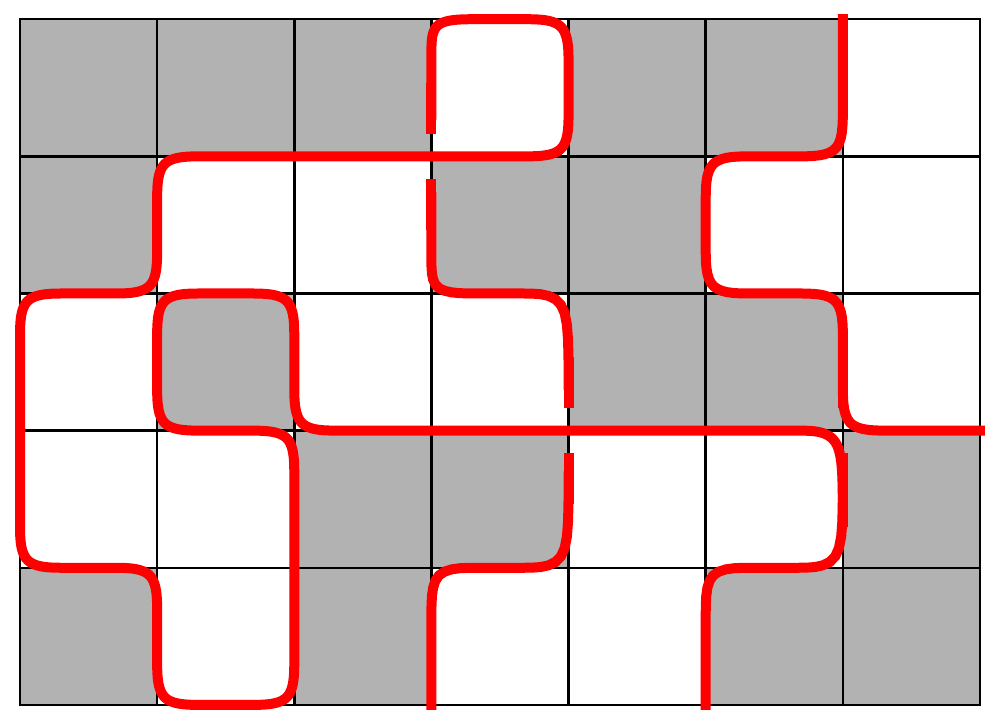}
\caption{Part of a configuration in the IPLC. In this model the loops (thick red lines) are cluster boundaries. Universal behaviour in the IPLC is expected to coincide with that in the CPLC.}
\label{squarelatticeloops}
\end{figure}

For pedagogical reasons, it will be useful to  introduce and discuss a second loop model before returning to the CPLC. Loops in the new model will no longer be completely packed, but nevertheless the universal properties will be the same. We  refer to this model as the incompletely-packed loop model with crossings, or IPLC.

To generate a configuration in the IPLC, we first colour the plaquettes of the square lattice black or white, giving a site percolation configuration on the square lattice formed by the plaquettes.  The loops in the IPLC are then cluster boundaries, as shown in Fig.~\ref{squarelatticeloops}. However the loop configuration is not uniquely determined by the cluster configuration: for each `doubly visited' node, where two cluster boundaries meet, we must choose how to connect them up. Allowing crossings, the three possible pairings are again those of  Fig.~\ref{decompnode} (but unlike in the CPLC we do not assign different weights to the different pairings).

The simplest choice for the Boltzmann weight is to give each percolation configuration the standard percolation probability $\q^\mathrm{B} (1-\q)^\mathrm{W}$, where $\mathrm{B}$ and $\mathrm{W}$ are the numbers of black and white faces. A given percolation configuration corresponds to  $3^N$ loop configurations, where $N$ is the number of doubly-visited nodes. Assigning them equal probability, the partition function for the IPLC is
\be
Z = \sum_{\text{configs}} \alpha^{N} \q^{\text{B}} (1-\q)^{\text{W}} , 
\ee
with $\alpha=1/3$. The parameter $\q$ here will play a similar role to the parameter $q$ in the CPLC.

\begin{figure}[b] 
\centering
\includegraphics[width=3.3in]{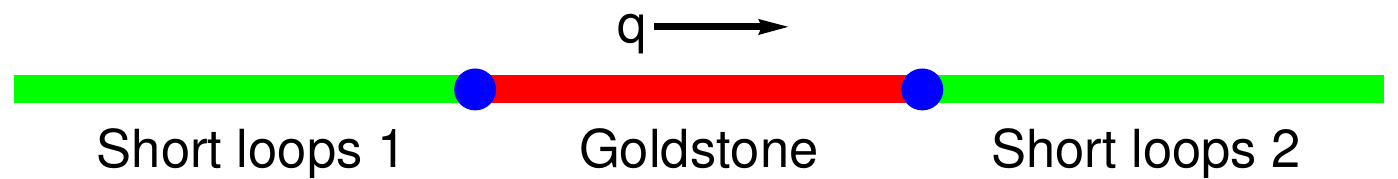}
\caption{Schematic phase diagram for the incompletely-packed loop model with crossings as a function of $\q$ (for fixed $n \sim 1$, $\alpha \sim 1/3$, $x \sim 1$).}
\label{model2phasediagram}
\end{figure}

The above partition function corresponds to a loop fugacity $n=1$. We will wish to generalize it to arbitrary $n$. We may also vary $\alpha$, and introduce a fugacity $x$ for the total \emph{length} of loops:
\be
Z = \sum_{\mathcal{C}} \alpha^{N}  \q^{\text{B}} (1-\q)^{\text{W}} x^\text{length} n^\text{no. loops}. 
\ee
The precise values of $\alpha$ and $x$ will not be important in what follows. When $n=1$, a length fugacity $x$ distinct from one corresponds to an Ising interaction of strength $J$ between colours of adjacent squares, with  $x=e^{-2J}$, and varying $\alpha$ introduces a four-site interaction. These interactions have no effect on the universal behaviour so long as they are weak.

Note that if we take our lattice to have the topology of the disk,  regarding the region outside the boundary as white for the purposes of drawing cluster boundaries, the IPLC shares with the CPLC the feature of having an edge loop at $\q=1$ but not at $\q=0$. There are clearly stable short-loop phases at small $\q$ and at $\q$ close to one, and field theory arguments lead us to expect a stable Goldstone phase near $\q=1/2$. The conjectured phase diagram, shown in Fig.~\ref{model2phasediagram}, is similar to a slice through the phase diagram of the CPLC at some nonzero value of $p$.  (If we had forbidden crossings, we would have obtained a phase diagram similar to the line $p=0$ in the CPLC.)

\section{Lattice field theories}
\label{lattice field theory section}

\subsection{Lattice field theory for IPLC}
\label{lattice gauge theory}

We begin with the IPLC, which permits a simple mapping to a $\mathbb{Z}_2$  lattice gauge theory coupled to matter fields. We will first write down this theory and then show that a  graphical expansion similar to the high temperature expansion of the Ising model or the Nienhuis $O(n)$ model \cite{domany, cardy book} provides the connection to the loop model. 

The required lattice field theory includes matter fields, which are real $n$-component vectors living on the sites $i$ of the square lattice,
\ba\label{S normalisation}
\vec S_i &= (S_i^1,\ldots, S_i^n), & \vec S_i^2& =n,
\end{align}
and gauge fields $\sigma_{ij}=\pm 1$ living on the links. The  partition function is:
\be\label{latticegaugetheory}
Z = \Tr \, \prod_{F} \biggl(
(1-\q) + \q \prod_{\<ij\>\in F} \sigma_{ij} \biggr)
 \prod_{\<ij\>} \lf 1 + x \sigma_{ij} \vec S_i. \vec S_j \ri. 
\ee
`$\Tr$' denotes sums and integrals over all degrees of freedom ($\sigma$ and $\vec S$), normalized so $\Tr 1 =1$, and $F$ denotes a face (square) of the lattice. The $\mathbb{Z}_2$ gauge symmetry of this model is
\ba
\vec S_i & \rightarrow \chi_i \vec S_i, &
\sigma_{ij} &\rightarrow \chi_i \chi_j \sigma_{ij} &
&(\text{for } \chi_i = \pm 1).
\end{align}
Above we have written the Boltzmann weight for the gauge field in a form suitable for the graphical expansion. Later we will rewrite it in a more conventional form.

The graphical expansion begins by expanding out the product over faces $F$ and representing the terms by a simple graphical rule (a face is coloured black if the $\q$ term is chosen, and white if the $1-\q$ term is chosen). This generates percolation configurations $\mathcal{P}$,
\be
Z = \sum_{\mathcal{P}} \q^{\text{B}} (1-\q)^{\text{W}} 
 \Tr  \lf \prod_{l \in \partial \mathcal{P} } \sigma_l \ri \prod_{\<ij\>} \lf 1 + x \sigma_{ij} \vec S_i. \vec S_j \ri,
\ee
where $\partial \mathcal{P}$ denotes the set of links lying on percolation cluster boundaries. 

Next we expand out the product over links. Only one term in this expansion survives after summing over $\sigma_{ij}$, namely that in which the factors of $\vec S_i . \vec S_j$ lie on the cluster boundaries:
\be
Z = \sum_{\mathcal{P}} \q^{\text{B}} (1-\q)^{\text{W}} x^\text{length}
 \Tr  \prod_{\<ij\> \in \partial \mathcal{P} }   S_i^a.  S_j^a.
\ee
`Length' refers to the total length of cluster boundaries. We have written the spin index $a$ explicitly in the inner product $\vec S_i .\vec S_j$ to emphasize that each link on a cluster boundary now carries an index $a=1,\ldots n$. Finally we perform the remaining integrals over the vectors $\vec S$, using
\ba \notag
\Tr S_i^a S_i^b &= \delta^{ab}, \\  \label{Straces}
\Tr S_i^a S_i^b S_i^c S_i^d &= \f{n}{n+2} \lf \delta^{ab}\delta^{cd} + \delta^{ac}\delta^{bd} + \delta^{ad}\delta^{bc}\ri.
\end{align}
The three terms in the second formula correspond to the three ways of connecting up the links in pairs at a node for which all four links are in $\partial \mathcal{P}$. Expanding out all such brackets gives $3^N$ terms, each associated with a loop configuration $\mathcal{C}$. For a given $\mathcal{C}$, each loop comes with a product of delta functions forcing the indices $a$ to be equal for all links on that loop. We may therefore think of each loop as carrying a `colour' index ranging from $1,\ldots, n$. Summing over the possible colours for each loop yields a fugacity $n$:
\be
Z = \sum_\mathcal{C} \alpha^N \q^B (1-\q)^W x^\text{length} n^\text{no. loops}.
\ee
The parameter $\alpha$ is $n/(n+2)$ as a consequence of Eq.~\ref{Straces}. It can be varied by exchanging the hard constraint $\vec S^2=n$ for a potential for $\vec S^2$.

\subsection{Correlation functions and the replica trick}
\label{correlation functions and replica}

\begin{figure}[b] 
\centering
\includegraphics[width=3.2in]{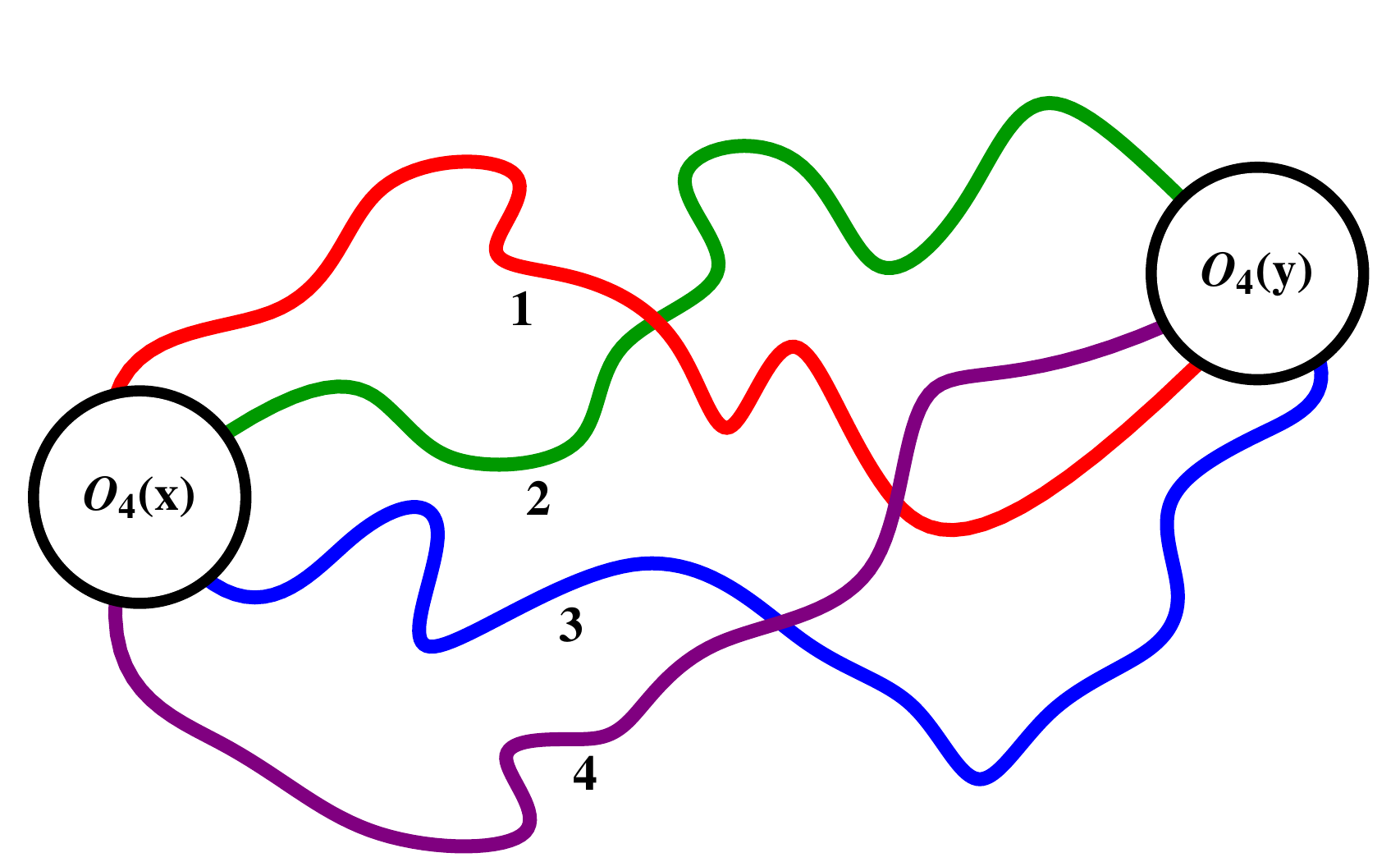}
\caption{The watermelon correlator $G_4(x,y)$.}
\label{g4correlator}
\end{figure}

We have seen that a graphical expansion of the lattice gauge theory (\ref{latticegaugetheory}) generates  the configurations of the loop model with the right weights. The correspondence also extends trivially to correlation functions. Switching temporarily to a continuum notation, the basic ones are the watermelon correlation functions $G_k(x,y)$, which give the probability that $k$ distinct strands of loop connect $x$ and $y$. For example, on the lattice $G_2$ can be taken to be the probability that two links lie on the same loop, and $G_4$ the probability that two nodes are connected by four strands. 

In the continuum, the correlator $G_k$ is the two-point function of the `$k$-leg' operator $O_k$:
\be
O_k (x) \propto S^1(x) S^2 (x) \ldots S^k (x).
\ee
$G_k$ vanishes for odd $k$, either by gauge invariance or equivalently because two regions cannot be joined by an odd number of strands (for example if two sites lie on the same loop they are joined by two strands). We can visualize the above operator, when inserted into a correlation function, as emitting $k$ strands with colour indices ranging from $1$ to $k$ (Fig~\ref{g4correlator}). 

We see that in order to write down the correlator $G_k$ we require the number $n$ of spin components to be at least $k$. This presents a problem if the model we wish to study has a loop fugacity $n<k$: for the model of most interest to us with $n=1$, it does not allow us to write down any of the above correlation functions! Fortunately, this problem can be resolved in two standard ways. 

Firstly, we can treat $n\rightarrow 1$ as a replica-like limit, so that
\be
G_k = \lim_{n\rightarrow 1} \< O_k (x) O_k (y) \>.
\ee
This trick may also be used for other values of the loop fugacity, integer or noninteger. For integer $n$ there is a more rigorous alternative, which is to use supersymmetry \cite{McKane, Parisi Sourlas SUSY for loops}. This allows us to increase the number of components of $\vec S$ without increasing the loop fugacity, by making some of the components fermionic. For our purposes the two approaches are equivalent, so for presentational simplicity we will use the replica language. The required supersymmetric construction is explained in Refs.~\cite{dense loops and supersymmetry, vortex paper}: in outline, $\vec S$ is replaced by a vector $\vec \Phi$ with both bosonic and fermionic components,
\ba \label{superspin}
\vec \Phi &= (\vec S, \vec \eta, \vec \xi), & 
\vec S^2 + 2 \vec \eta. \vec \xi = n,
\end{align}
where $\vec S$ has $n+2m$ components and $\vec \eta$, $\vec \xi$ each have $m$ anticommuting components. The loop fugacity determines $n$, but $m$ is arbitrary -- supersymmetry ensures that the partition function and its loop representation do not depend on it. Thus $m$ may be chosen large enough for any desired correlator to be constructed.

\subsection{$\mathbb{Z}_2$ vortices and $\mathbb{Z}_2$ fluxes}
\label{absence of gauge fluctuations}

\begin{figure}[b] 
\centering
\includegraphics[width=2.7in]{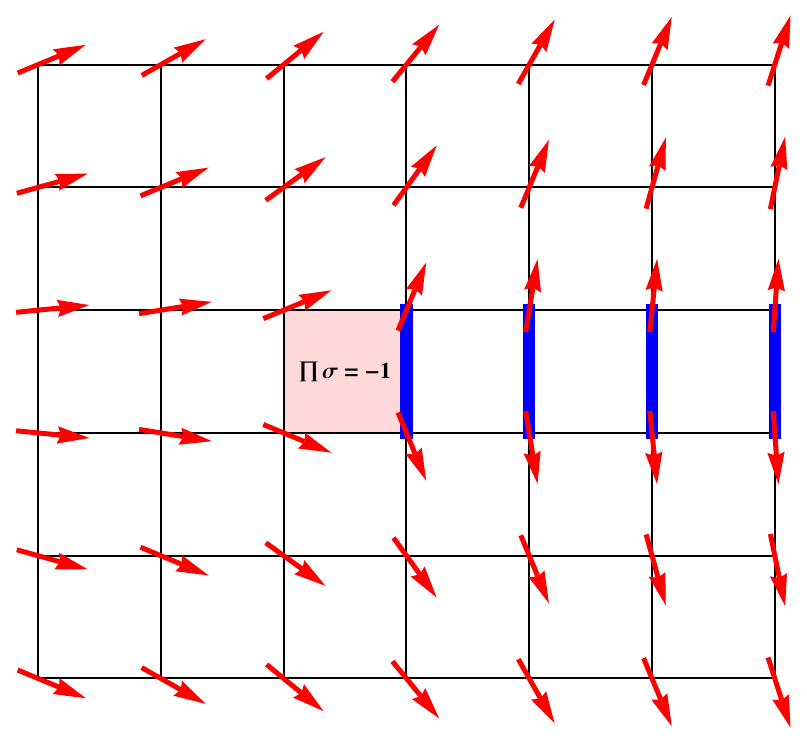}
\caption{Plaquettes with gauge flux $\prod\sigma = -1$ (shaded in pink) are endpoints of strings of links with $\sigma=-1$ (marked in bold/blue), across which $\vec S$ changes sign. In terms of the nematic order parameter, which is obtained by identifying $\vec S$ with $-\vec S$ and which lives on $\rp^{n-1}$, these plaquettes are vortices. (This is of course a caricature, neglecting fluctuations.)}
\label{vortexgraphic}
\end{figure}

The local gauge-invariant degree of freedom in the lattice gauge theory (\ref{latticegaugetheory}) is a nematic vector, obtained by identifying $\vec S$ with $-\vec S$. It can be encoded in a real symmetric matrix,
\ba
Q^{ab} & = S^a S^b - \delta^{ab}, &
\tr Q & = 0,
\end{align}
which will be the relevant degree of freedom on long length scales. $Q$ lives on real projective space, $\rp^{n-1}$: since this manifold has nontrivial fundamental group \cite{mermin defect review}, the $Q$ configuration can have vortex defects which we now discuss.

$\rp^{1}$ is equivalent to the circle, so at the special value $n=2$ vortices are standard XY vortices and are characterized by an integer topological charge. However $\pi_1(\rp^{n-1})$ = $\mathbb{Z}_2$ when $n>2$, so in general the vortex charge is defined only modulo two.  In the replica limit ---  which requires analytic continuation of formulae defined for arbitrarily large $n$ to $n<2$ --- the vortices should again be viewed as $\mathbb{Z}_2$ vortices. (This may be clearer in the supersymmetric formulation of the $n=1$ loop model, Eq.~\ref{superspin}, where the bosonic part of the superspin lives on $\rp^{2m}$ for $m\geq 1$, which has fundamental group $\mathbb{Z}_2$.)

For a more concrete picture we return to $\vec S$ and $\sigma$. Let $\sigma_l = +1$ on all links $l$, except for a semi-infinite string of parallel links ending at a plaquette $F$ (see Fig.~\ref{vortexgraphic}). The flux $\prod \sigma$ is then $-1$ only on $F$. The spin configurations  which maximize the Boltzmann weight vary smoothly except at the string, across which $\vec S$ changes sign. This indicates the presence of a $\mathbb{Z}_2$ vortex located at $F$. (This connection between vortices in nematics and $\mathbb{Z}_2$ fluxes is standard \cite{lammert et al}.)

Since vortices are associated with plaquettes of nontrivial flux, we can assign a vortex number to each plaquette which is $+1$ if $\prod \sigma =-1$ and $0$ if $\prod \sigma =1$. The gauge field part of the Boltzmann weight can then be written in terms of the number $N_v$ of vortices:
\be\label{vortex fugacity}
\prod_{F} \lf (1-\q) + \q \prod_{\<ij\>\in F} \sigma_{ij} \ri = (1-2\q)^{N_v}.
\ee
We see that the factor $(1-2\q)$ is simply a fugacity for vortices, and that the exchange $\q\leftrightarrow 1-\q$ corresponds to changing the sign of the vortex fugacity.

This sign distinguishes the two short-loop phases in Fig.~\ref{model2phasediagram} from each other.  In Ref.~\cite{Fu Kane}, $\mathbb{Z}_2$ vortices play an analogous role in a sigma model for localisation, with the sign of the vortex fugacity distinguishing two insulating phases. (The vortex fugacity is also important for the critical behaviour \cite{Fu Kane} --- see Sec.~\ref{RG equations with vortices}.)

Let us rewrite the Boltzmann weight for $\sigma$ in the conventional form for $\mathbb{Z}_2$ gauge theory. In the absence of a boundary, Eq.~\ref{latticegaugetheory} may be written
\be \label{rewrite sigma weight}
Z \propto \Tr \exp \biggl( \kappa \sum_F \prod_{\< ij\> \in F} \sigma_{ij}  \biggr)
\prod_{\<ij\>} \lf 1 + x \, \sigma_{ij} \vec S_i. \vec S_j \ri,
\ee
where the gauge field stiffness is
\be
\label{gauge field stiffness}
\kappa = \f{1}{2}\ln \f{1}{| 1-2 \q |}.
\ee
In the presence of a boundary, denoted $\partial$, the Boltzmann weight acquires an additional term when $\q>1/2$:
\be\label{boundary term}
\prod_{\<ij\>\in \partial} \sigma_{ij}.
\ee
This term effects the sign change in the vortex fugacity. It is equal to $(-)^{N_\text{strings}}$, where $N_\text{strings}$ is the number of $\sigma = -1$ strings which terminate on the boundary. Since this number is equal to the number of vortices in the interior modulo 2, $(-)^{N_\text{strings}}=(-)^{N_v}$.

We see that the sign of the vortex fugacity does not affect bulk properties. Instead it determines the presence or absence of an edge loop.

Finally, consider the point $\q=1/2$. The vortex fugacity vanishes here (Eq.~\ref{vortex fugacity}); however, the universal properties of this point do not differ from those in the rest of the Goldstone phase (Fig.~\ref{model2phasediagram}). This is because vortices are anyway RG irrelevant in that phase. We discuss this in the next section in terms of a sigma model for $Q$.

The suppression of vortices (either microscopically or in the infrared) means that in the Goldstone phase this sigma model has a correspondence with the simpler $O(n)$ sigma model. In the IPLC at $\q=1/2$ this correspondence holds microscopically, since the gauge field stiffness diverges there (\ref{gauge field stiffness}). This enforces $\prod \sigma=1$ for every face $F$, giving $\sigma_{ij}= \chi_i \chi_j$ (so long as the lattice lives on a simply-connected manifold\footnote{Otherwise there can be flux through holes or handles of the manifold. In the graphical expansion, the sum over the resulting flux sectors is responsible for killing loop configurations which do not correspond to percolation configurations.})
\be\label{partition function with chi}
Z\propto \sum_{\{\chi\}} \Tr_S \prod_{\<ij\>} \lf 1 + x \, (\chi_i \vec S_i).(\chi_j \vec S_j) \ri. 
\ee
Changing variables to $\vec S' = \chi \vec S$ eliminates the gauge degrees of freedom from the Boltzmann weight, leaving a lattice $O(n)$ model. (Non-gauge-invariant correlators pick up factors of $\chi$, ensuring that they vanish\footnote{However, at this special value of $\q$ we can define $k$-leg operators for odd $k$ by taking $O_k=S'^1\ldots S'^k$.} on summing over $\chi$.) In the regime we consider, i.e. at sufficiently large $x$, this $O(n)$ model is expected to be described by the $O(n)$ sigma model in its Goldstone phase.

\subsection{Continuum description}
\label{continuum description}

The naive continuum description of the lattice field theory (\ref{latticegaugetheory}) is a sigma model for $Q$,
\be
\label{introduction of sigma model}
\mathcal{L} =  \f{K}{4} \tr \, (\nabla Q)^2,
\ee
together with the constraint on $Q$  following from its definition in terms of $\vec S$.  To conform with convention we use the normalisation $\vec S^2 =1$ in the continuum (rather than Eq.~\ref{S normalisation}); then  $Q^{ab} = S^a S^b - \f{1}{n}  \delta^{a b}$.

The fugacity for vortices is hidden in the ultraviolet regularisation of the sigma model (\ref{introduction of sigma model}). We will restore this parameter explicitly when we consider RG in the vicinity of the critical point (Sec.~\ref{RG equations with vortices}), where vortices are crucial.

However, since the classical free energy of a pair of vortex defects is proportional to the stiffness $K$, they are suppressed at large $K$ (as in the XY model at large stiffness). In the Goldstone phase, which we now discuss, $K$ flows to large values under coarse-graining, and vortices are an irrelevant perturbation. 

Non-singular (i.e. vortex-free) configurations of $Q$ are equivalent to non-singular configurations of $\vec S$ (on a manifold of trivial topology, and up to a global sign ambiguity). Thus for a perturbative treatment at large $K$, the $\rp^{n-1}$ sigma model can be replaced with the more familiar $O(n)$ sigma model,  argued previously \cite{dense loops and supersymmetry} to apply to the CPLC at $q=1/2$:
\ba
\label{introduction of O(n) sigma model}
\mathcal{L} &=  \f{K}{2}  (\nabla \vec S)^2, & \vec S^2 & =1.
\end{align}
The perturbative beta function for $K$ changes sign at $n=2$ \cite{dense loops and supersymmetry, Polyakov}:
\be
\label{basic RG equation}
\f{\dd K}{\dd \ln L} = \f{2-n}{2\pi} \lf 1 + \f{1}{2 \pi K} + \ldots \ri
\ee
When $n>2$, the stiffness flows to zero under RG and the sigma model has only a disordered phase; thus the loop model is not expected to be critical. At $n=2$ the sigma model is the XY model, and we have in addition the quasi-long-range-ordered phase in which $K$ does not flow. By the Mermin Wagner theorem, these are the only possibilities when $n\geq 2$.

However in the replica limit the Mermin Wagner theorem does not apply \cite{dense loops and supersymmetry}, and Eq.~\ref{basic RG equation} shows that for $n<2$ the stiffness $K$ flows to infinity in the infrared. This is the Goldstone phase. At the infra-red fixed point the $n-1$ Goldstone modes are free fields, so the central charge is $c= n-1$ \cite{dense loops and supersymmetry}.

We now discuss the extraction of a continuum description for the CPLC, the appearance of a nontrivial vortex fugacity in that model, and the extra symmetry on the boundaries of the phase diagram.

\subsection{Lattice field theory for the CPLC}
\label{Lattice field theory for the CPLC}

The CPLC can again be mapped to a lattice spin model with a $\mathbb{Z}_2$ gauge symmetry, though with a less conventional form, and as long as we are in the interior of the phase diagram (Fig.~\ref{phasediagram}) the continuum description is again the $\rp^{n-1}$ model. The following construction is similar to that described in Ref.~\cite{short loop paper}  so we give only an outline. 

We again introduce fixed-length spins $\vec S_l$ ($\vec S_l^2 = n$) but for the CPLC they live on the links $l$ of the square lattice. The Boltzmann weight involves four-spin interactions between the spins surrounding each node $i$:
\be
\label{ZforCPLC}
Z = \Tr \, \exp \lf - \sum_{\text{nodes }i} E_i \ri. 
\ee
In order to define $E_i$, we denote the links surrounding $i$ by $1, 2,3, 4$, with the weight $p$ pairing being  $1$ with $3$ and $2$ with $4$, and the weight $(1-p)q$ pairing being $1$ with $2$ and $3$ with $4$:
\ba\notag
\exp\lf - E_i \ri  = & \, p (\vec S_1. \vec S_3) (\vec S_2. \vec S_4)  + 
(1-p)q (\vec S_1. \vec S_2)(\vec S_3. \vec S_4) \\ 
\notag
& + (1-p)(1-q) (\vec S_1. \vec S_4)(\vec S_2. \vec S_3).
\end{align}
These three terms are in correspondence with the three possibilities in Fig.~\ref{decompnode}, and a graphical expansion of the above partition function gives the sum over loop configurations defining the CPLC (Eq.~\ref{ZCPLC}). 

The above Boltzmann weight again has a $\mathbb{Z}_2$ gauge symmetry: on changing the sign of the spin on the link $ij$, both $e^{-E_i}$ and $e^{-E_j}$ change sign but the overall Boltzmann weight is unchanged. The naive continuum limit is again the $\rp^{n-1}$ sigma model described above, and the universal properties are expected to be identical with those of the IPLC. In Sec.~\ref{vortex fugacity in CPLC} we will argue that the effective fugacity for $\mathbb{Z}_2$ vortices changes sign on the line $q=1/2$, just as for the IPLC.

The lattice field theory (\ref{ZforCPLC}) and the continuum limit (\ref{introduction of sigma model}) require a couple of comments. Firstly, the $\rp^{n-1}$ description does not apply on the boundaries of the phase diagram: here the model has a higher symmetry which is not taken into account in (\ref{ZforCPLC}) --- see next section. Secondly, note that $\exp(-E_i)$ is not always positive. This is not an obstacle to taking the continuum limit --- in fact it is an important feature of the problem (note that complex lattice actions are not necessarily pathological, and are the norm in quantum problems). We have seen that different signs for the vortex fugacity distinguish the two short-loop phases in the IPLC, and we will see that the same is true for the CPLC. (The possibility of non-positive Boltzmann weights can important even in the Goldstone phase -- Sec.~\ref{spanning number section}).

In the model without crossings, discussed below, it is again crucial that the lattice Boltzmann weight is not positive since the continuum Lagrangian contains an imaginary $\theta$-term \cite{candu et al}. We now briefly review this formulation and discuss its implications for the phase diagram of the CPLC.

\subsection{Phase diagram boundaries and $\cp^{n-1}$}
\label{boundaries of phase diagram}

As mentioned in Sec.~\ref{CPLC}, the CPLC has an additional symmetry \cite{read saleur, candu et al} on the boundaries of the phase diagram (when $p=0$, or $q=0$, or $q=1$). On each of the three boundaries of the phase diagram, the links of the lattice can be assigned fixed orientations, with two incoming and two outgoing links at each node, such that the allowed pairings are always between an incoming and an outgoing link. This orients all the loops. The necessary choice of link orientations differs for each of the components of the boundary, as mentioned in Sec.~\ref{CPLC}.

Then instead of constructing a lattice field theory using real spins $\vec S$, we may take complex spins $\vec z$. The inner products $\vec S_l. \vec S_{l'}$ in the Boltzmann weight for a given node are replaced with ${\vec z_l}^{\,\dag}. \vec z_{l'}$, where $l$ is the outgoing link and $l'$ the incoming link. One way of understanding the appearance of complex fields is to view the loops as worldlines of quantum particles in $1+1$ dimensional spacetime: the fact that we are now dealing with oriented worldlines means that these particles carry a $U(1)$ charge.

The $SO(n)$ global and $\mathbb{Z}_2$ gauge symmetry in the interior of the CPLC phase diagram are promoted to an $SU(n)$ global and $U(1)$ gauge symmetry on its boundary. The appropriate field theory is a sigma model for a field on \emph{complex} projective space ($\cp^{n-1}$), with a $\theta$ term \cite{read saleur, candu et al}. Complex projective space is the manifold of unit vectors $\vec z$ modulo the gauge equivalence $\vec z \sim e^{i\phi} \vec z$; again a non-redundant parametrisation is provided by a traceless matrix $\tilde Q$, which is now Hermitian rather than real symmetric ($\tilde Q^{ab} = z^a z^{*b} - \delta^{ab}/n$). The Lagrangian for the $\cp^{n-1}$ sigma model is
\be
\mathcal{L}_{\mathbb{CP}^{n-1}}= \f{K}{4} \tr \, (\nabla \tilde Q)^2 +  \f{\theta}{2 \pi}  \epsilon_{\mu \nu }
\tr \tilde Q \nabla_\mu \tilde Q \nabla_\nu \tilde Q.
\ee
The field $\tilde Q$ can sustain skyrmion textures, since $\pi_2(\cp^{n-1})=\mathbb{Z}$. The $\theta$ term weights skyrmions by a factor $e^{i\theta}$ and antiskyrmions by $e^{-i\theta}$. In the case $n=2$,   the above field theory is equivalent to the $O(3)$ sigma model (the $O(3)$ spin is equal to $\tr \vec \sigma \,\tilde Q$, where $\vec \sigma$ is a vector of Pauli matrices); for a pedagogical discussion of the $\theta$ term in this case see for example Ref. \cite{fradkin book}.

Bulk properties of the $\cp^{n-1}$ sigma model depend on $\theta$ only modulo $2\pi$. For $n\leq2$, there is a critical point at $\theta=\pi$ $\text{mod} \, 2\pi$; other values of $\theta$ are massive (flowing under RG to $\theta=0\text{ mod } 2\pi$, $K=0$). The critical point at  $\theta=\pi$ is the critical point of the loop model at  $p=0$, $q=1/2$  \cite{read saleur, candu et al}. Close to this point the bare value of $\theta$ varies as
\be
(\theta-\pi) \propto (q-1/2).
\ee

For our present purposes the $\cp^{n-1}$ description tells us two things. Firstly, it implies that the left- and right-hand boundaries of the phase diagram Fig.~\ref{phasediagram}, which correspond to the Manhattan lattice loop model, are localized for all $p$ as previously expected \cite{beamond cardy chalker, beamond owczarek cardy}. Here $\theta$ can be shown\footnote{One way to show this explicitly is to map an anisotropic version of the loop model to a quantum spin chain with next-nearest-neighbour couplings, which can be coarse-grained using standard techniques.} to be equal to zero $\mathrm{mod}\,\,2\pi$, so that the sigma model is in the disordered phase. However the correlation length $\xi$, and the typical loop size, diverge exponentially as $p\rightarrow 1$,
\ba\notag
\xi & \sim (1-p)^{-2} e^{\text{const.}/(1-p)} & (\text{when } q &=0,1).
\end{align} 
This follows from the beta function for the $\cp^{n-1}$ model \cite{CPn-1 beta function} and the fact that the bare stiffness is of order $(1-p)^{-1}$. It is the behaviour of $\xi$ for non-critical localisation in class C  \cite{quasiparticle localisation in high Tc}, to which the Manhattan lattice loop model is related \cite{beamond cardy chalker}.

Secondly, the $\cp^{n-1}$ description of the model without crossings  gives a way of seeing that the vortex fugacity in the $\rp^{n-1}$ description of the CPLC changes sign on the central line $q=1/2$, just as it does at the central point $\q=1/2$ in the IPLC. We now discuss this issue.

\subsection{Vortex fugacity and the $\theta$-term for $\mathbb{CP}^{n-1}$}
\label{vortex fugacity in CPLC}

A natural way to approach the field theory for the CPLC, at least when the weight $p$ for crossings is small, is by perturbing the field theory for the model without crossings. In the $\cp^{n-1}$ language, making the weight $p$ for crossings nonzero corresponds to adding a small mass for the imaginary part of $\tilde Q$, i.e. $\delta \mathcal{L} \propto - p \tr \, (\text{Im} \, \tilde Q)^2$  \cite{vortex paper}. This anisotropy favours real $\tilde Q$ and leads in the infrared to an order parameter living on $\rp^{n-1}$. 

If we simply set $\tilde Q$ to be real, the kinetic term for $\mathcal{L}_{\cp^{n-1}}$ becomes that of $\mathcal{L}_{\rp^{n-1}}$, and the $\theta$ term vanishes. Thus we regain the $\rp^{n-1}$ sigma model Lagrangian (\ref{introduction of sigma model}) for the loop model with crossings.

However, this does not mean the $\theta$ term plays no role: it vanishes only if we neglect $\rp^{n-1}$ vortices. In the presence of a vortex, we must allow the imaginary part of $\tilde Q$ to become nonzero in the vortex core, in order to retain continuity of $\tilde Q$ and the sigma model constraint. There is then a contribution from the $\theta$ term there. This mechanism also pertains to quantum magnets described by anisotropic sigma models in both $1+1$ and $2+1$ dimensions \cite{Affleck mass generation, Senthil Fisher competing orders}. 

The vortex core corresponds either to a half-skyrmion or to an anti-half-skyrmion, depending on the sign  of the imaginary components of $\tilde Q$. (This can easily be visualised for $n=2$, when the perturbed $\cp^1$ model is equivalent to the $O(3)$ sigma model with easy-plane anisotropy: a half-skyrmion corresponds to a vortex in the easy plane, with the $O(3)$ spin perpendicular to the easy plane in the core.) The effective vortex fugacity $V$ is obtained by summing over both possibilities, half-skyrmion and anti-half-skyrmion, leading to
\ba\notag
V \propto e^{i \theta/2}+ 
e^{-i \theta/2}.
\end{align}
Since at $p=0$ we have $(\theta-\pi) \sim (q-1/2)$,  the vortex fugacity in the $\rp^{n-1}$ description of the CPLC changes sign at $q=1/2$, just as the vortex fugacity changes sign at $\q=1/2$ in the IPLC.

\section{The Goldstone phase}
\label{Goldstone phase}

The Goldstone phase shows subtle universal behaviour, different from that seen in loop models without crossings, which can however be understood in detail. Within this phase we can work with the $O(n)$ sigma model  \cite{dense loops and supersymmetry}, as discussed in Sec.~\ref{continuum description}. 

For most purposes the Lagrangian (\ref{introduction of O(n) sigma model}) will be sufficient, but to calculate the length distribution in Sec.~\ref{length distribution} we will need to add a small perturbation, $\gamma$, which breaks the symmetry from $O(n)$ to $O(n-1)\times\mathbb{Z}_2$. Writing 
\be  \label{S split}
\vec S = (S^1, \vec S_\perp),
\ee
where $\vec S_\perp$ is an $(n-1)$-component vector, we take
\ba
\label{perturbed lagrangian}
\mathcal{L}  &= \f{K}{2} \lf ( \nabla \vec S )^2 + \gamma \, S_\perp^2 \ri, &
\vec S^2 & = 1.
\end{align}
We briefly recall one-loop RG results for this model \cite{Polyakov, Pelcovits and Nelson}, which may be obtained easily using the background field method. $S(x)$ is decomposed into a slowly-varying field $\tilde S(x)$ and rapidly varying fluctuations $\phi(x)$:
\be
\vec S (x) = \vec {\tilde S}(x)  \sqrt{1 - \phi(x) ^2} + \sum_{\beta=1}^{n-1} \phi_\beta (x)  \vec e_\beta (x) .
\ee
The $e_\beta(x)$, $\beta=1,\ldots, n-1$, are a set of vectors orthogonal to ${\tilde S}(x)$ (there is a gauge freedom in this choice). If the initial UV cutoff is $\Lambda$, so that $S$ involves modes with wavenumber $|k| < \Lambda$, then the modes in $\tilde S$ are limited to $|k|< \tilde \Lambda$ for the new cutoff $\tilde \Lambda < \Lambda$, and $\phi$ contains modes in the momentum shell $|k| \in (\tilde \Lambda, \Lambda)$. Integrating $\phi$ out, and working to leading order in $K^{-1}$ and $\gamma$, we obtain the RG equations
\ba
\label{K gamma RG equations}
\f{\dd K}{\dd \tau} & = \f{2-n}{2\pi}, &
\f{\dd \gamma}{\dd \tau} & = \lf 2 - \f{1}{\pi K} \ri \gamma.
\end{align}
Here $\tau$ is the RG time: after time $\tau$, the new cutoff is $e^{-\tau} \Lambda$. Again, the important point is that $K$ flows to large values in the infrared if $n<2$ \cite{dense loops and supersymmetry}. 

Note that in two dimensions the higher-order anisotropies (higher powers of $S_\perp^2$) are as relevant as $S_\perp^2$ at tree level, but less relevant at one-loop order. They will be important for the polymer phase diagram discussed in Sec.~\ref{polymer section}.

We now calculate a range of observables both analytically and numerically; details of the numerical procedure are given in Sec.~\ref{Numerical methods}. As we discuss in Sec.~\ref{polymer section}, incompatible hypotheses about the universal behaviour at the point $n=1$, $p=1/3$, $q=1/2$ have previously been put forward, which is one reason for making a careful comparison of numerics and theory in the Goldstone phase.

\subsection{Correlation functions}
\label{correlation function section}

\begin{figure}[t] 
\centering
\includegraphics[width=3.3in]{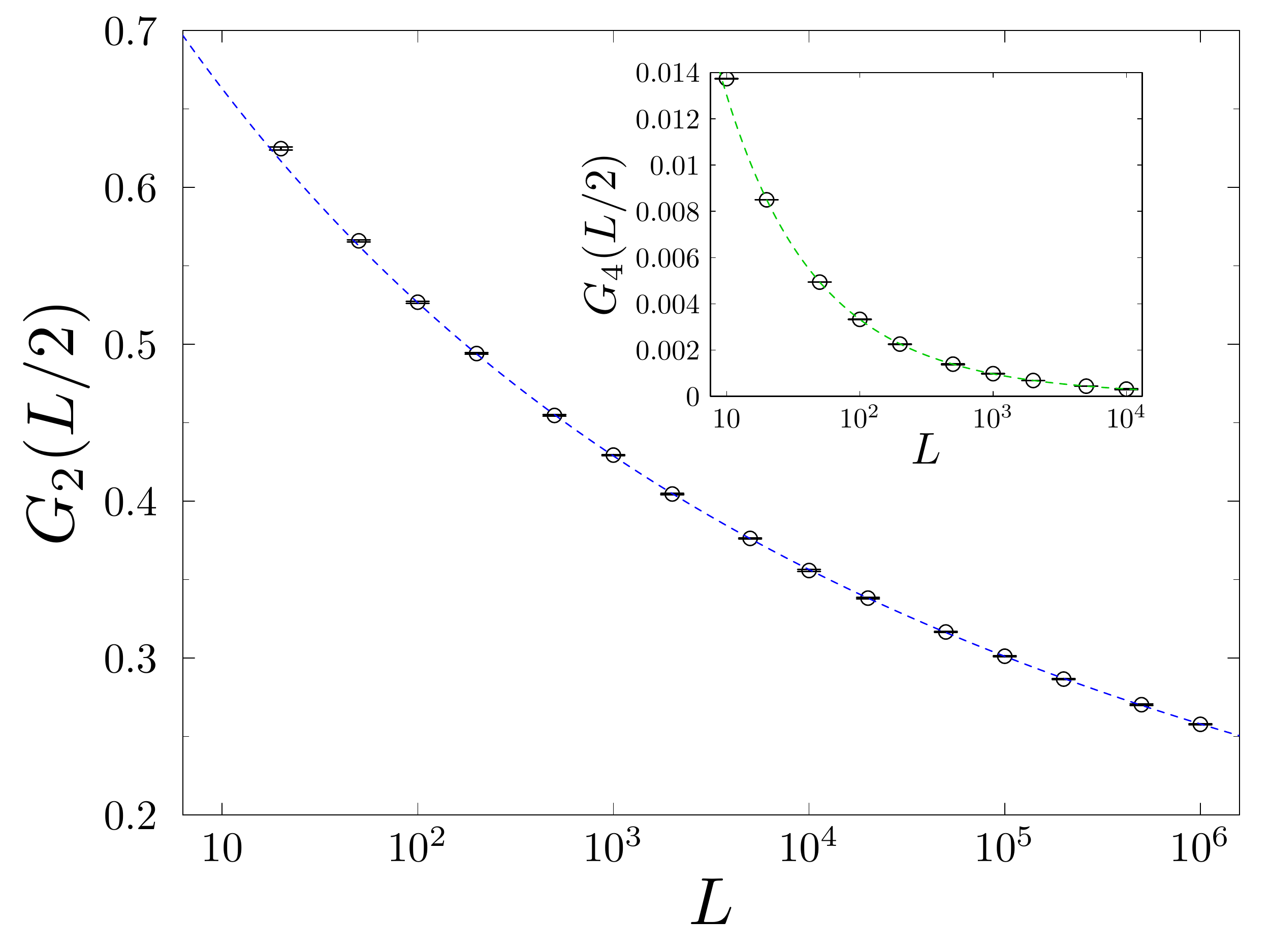}
\caption{Two- and four-leg watermelon correlation functions $G_2$ and $G_4$ in the Goldstone phase. The fits are to the form $G_k = \tilde C_k(\ln L/r_k)^{\tilde \alpha_k}$ ($k=2,4$), see text.}
\label{extphasecorrelators}
\end{figure}

The watermelon correlation function $G_k(r)$ may be expressed as the two-point function of the operator $S^1\ldots S^k(x)$ (Sec.~\ref{correlation functions and replica}). Including the UV cutoff $\Lambda$ and the sigma model stiffness $K$ explicitly in the argument of $G_k$, a simple calculation following \cite{Polyakov} gives  
\be
\label{correlator intermediate}
G_k (r \Lambda, K) = \f{G_k \lf 1, K + \f{2-n}{2\pi} \ln \Lambda r \ri}{\lf 1 + \f{2-n}{2\pi K} \ln \Lambda r \ri^{\alpha_k} },
\ee
where the exponent in the denominator depends on $k$ and on the loop fugacity $n$:
\be
\label{alpha}
\alpha_k = \f{k \, (k + n-2)}{2-n}.
\ee
The correlation function in the numerator of (\ref{correlator intermediate}), which is evaluated at a separation of the order of the new UV cutoff, tends to a constant for large $r$. Thus the asymptotic behaviour of the watermelon correlation functions $G_k(r)$ is given by a universal power of $\log r$:
\ba
\label{correlator final}
G_k (r ) & = \f{C_k}{\lf \ln r/r_0 \ri^{\alpha_k} }, &
r_0 = \Lambda^{-1} e^{-\f{2\pi K}{ 2-n}},
\end{align}
with nonuniversal $C_k$, $r_0$.

It is interesting to note that although the stiffness $K$ flows to infinity in the Goldstone phase --  which we would usually think of as implying long range order --- \emph{all} the correlation functions $G_{2l}$ decay to zero at large distances for $n>0$. Correlators $G_k$ with odd $k$ have no meaning in the $\rp^{n-1}$ sigma model or in the CPLC, but can be defined in loop models described by the $O(n)$ sigma model without a $\mathbb{Z}_2$ gauge symmetry; such models allow operators which insert dangling ends. For $n=1$, $G_1$ tends to a constant, indicating that the entropic force between the ends of an \emph{open} strand inserted into such a soup of closed loops  vanishes at large separations.

We may contrast this with the case $n=0$, which describes the universality class of the dense polymer with crossings \cite{dense loops and supersymmetry} --- a single loop whose length is comparable with the number of lattice sites. Here, $G_2(x)$ is a constant at large separations, in consequence of this fact. On the other hand $G_1(x)$ has a \emph{negative} exponent $\alpha_1$, indicating that if the polymer is an open strand the two ends suffer a weak entropic repulsion.

We might have expected the logarithmic form of (\ref{correlator final}) to prevent us from seeing the universal exponents $\alpha_k$ numerically, but this is not the case. Fig.~\ref{extphasecorrelators} shows $G_2(L/2)$ and $G_4(L/2)$ for $L\times L$ systems with periodic boundary conditions, with $L$ ranging up to $L=10^6$ for $G_2$ and $L=10^4$ for $G_4$ (see Sec.~\ref{Numerical methods} for further details). Simulations are at $p=q=1/2$. We fit $G_2$ and $G_4$ to the form $G_k = \tilde C_k(\ln L/r_k)^{-\tilde \alpha_k}$, leading to exponents consistent with (\ref{alpha}):
\ba
\tilde \alpha_2 & = 1.9 (1)  & \tilde \alpha_4 &  = 12.5 (10).
\end{align}
We have $\ln r_2 = - 15.4 (14)$, $\ln r_4 = -18 (2)$, consistent with the fact that  $r_0$ is shared between different $G_k$ in Eq.~\ref{correlator final}.

\subsection{Spanning number}
\label{spanning number section}

The logarithmic RG flow of the sigma model stiffness $K$ can be seen empirically: this stiffness is directly related to the mean `spanning number' for an $L\times L$ cylinder on which curves are allowed to terminate on the boundary. This is the number $n_s$ of curves which traverse the cylinder from one boundary to the other. Note that $n_s$ must be even if $L$ is even and odd if $L$ is odd.

To calculate $n_s$, the correspondence of Sec.~\ref{Lattice field theory for the CPLC} between the loop model and a spin model must be extended to the case with dangling boundary links. We simply take the spins on the dangling links to be fixed, with 
\ba
\notag
\f{\vec S_\text{top}}{\sqrt n} &= (\cos\theta, \sin\theta, 0, \ldots, 0), & \f{\vec S_\text{bottom}}{\sqrt n}  & = (1, 0, \ldots, 0)
\end{align}
on the top and bottom boundaries (above we temporarily revert to the lattice normalization of $\vec S$). The graphical expansion then goes through as before, except that spanning curves acquire an additional weight $\cos \theta$.\footnote{Loops in the interior and curves with both ends on the same boundary both retain fugacity $n$  --- for example a curve with both ends on the upper boundary can either be of colour $a=1$, in which case it has weight $n \cos^2 \theta$, or of colour $a=2$, when it has weight $n\sin^2\theta$, and the sum gives $n$. Spanning curves on the other hand must be of colour $a=1$, and have weight $n \cos\theta$.}

\begin{figure}[t] 
\centering
\includegraphics[width=3.3in]{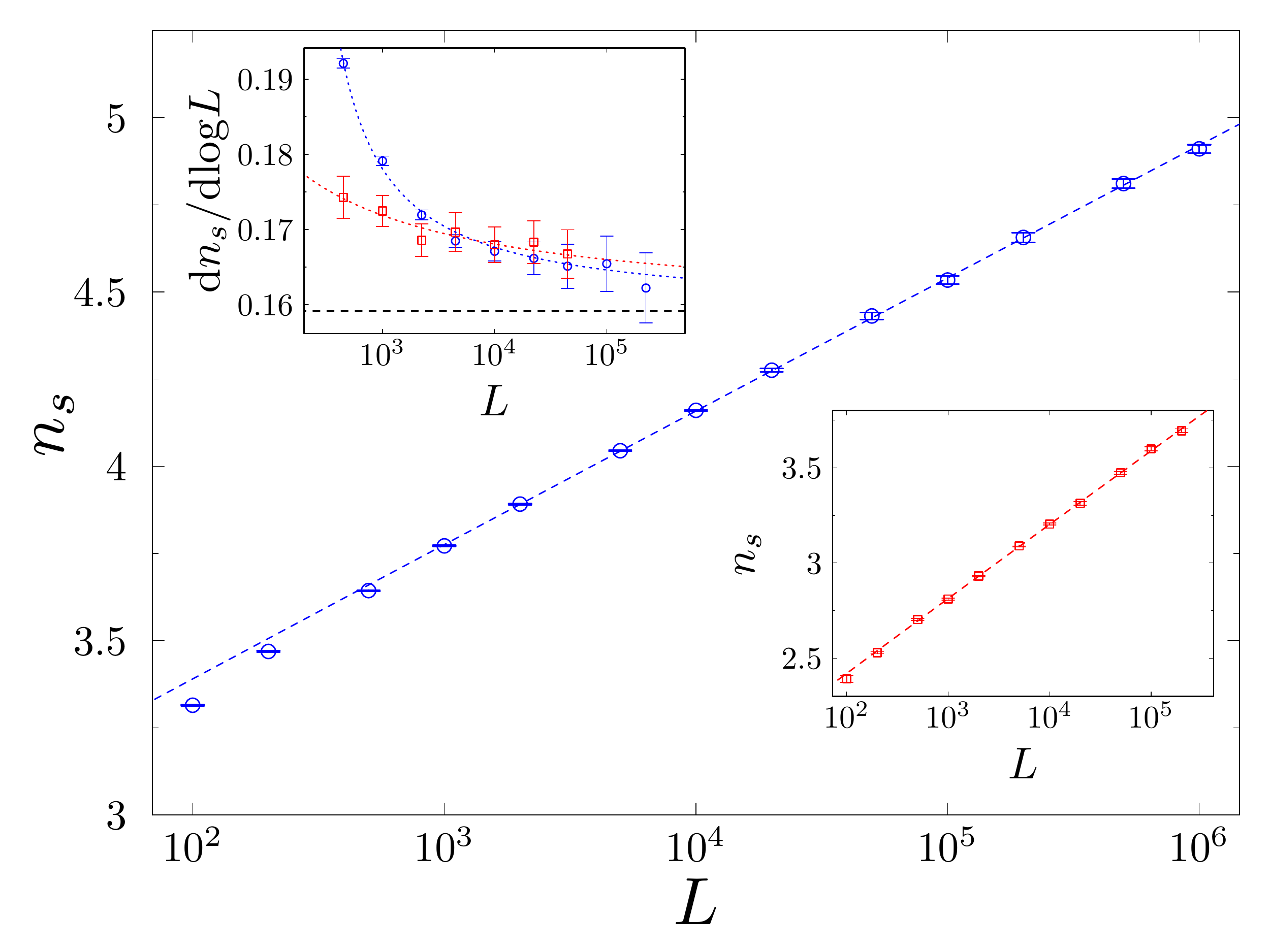}
\caption{The logarithmic increase of the mean spanning number $\<n_s\>$ with system size in the Goldstone phase. Main panel: $p=1/2$, $q=1/2$; lower inset: $p=1/3$, $q=1/2$. Fits are to $\f{1}{2\pi} (\ln L/L_0 + \ln \ln L/L_0)$ with $\ln L_0 \simeq -13.78, \, -8.06$ for $p=1/2$ and $p=1/3$ respectively. Upper inset shows the numerical estimates for the slopes $\dd \<n_s\> / \dd \ln L$ plotted against $\ln L$ -- they are expected to converge to $1/2\pi$ (horizontal line).}
\label{spanning number fig}
\end{figure}

\begin{figure}[t] 
\centering
\includegraphics[width=3.3in]{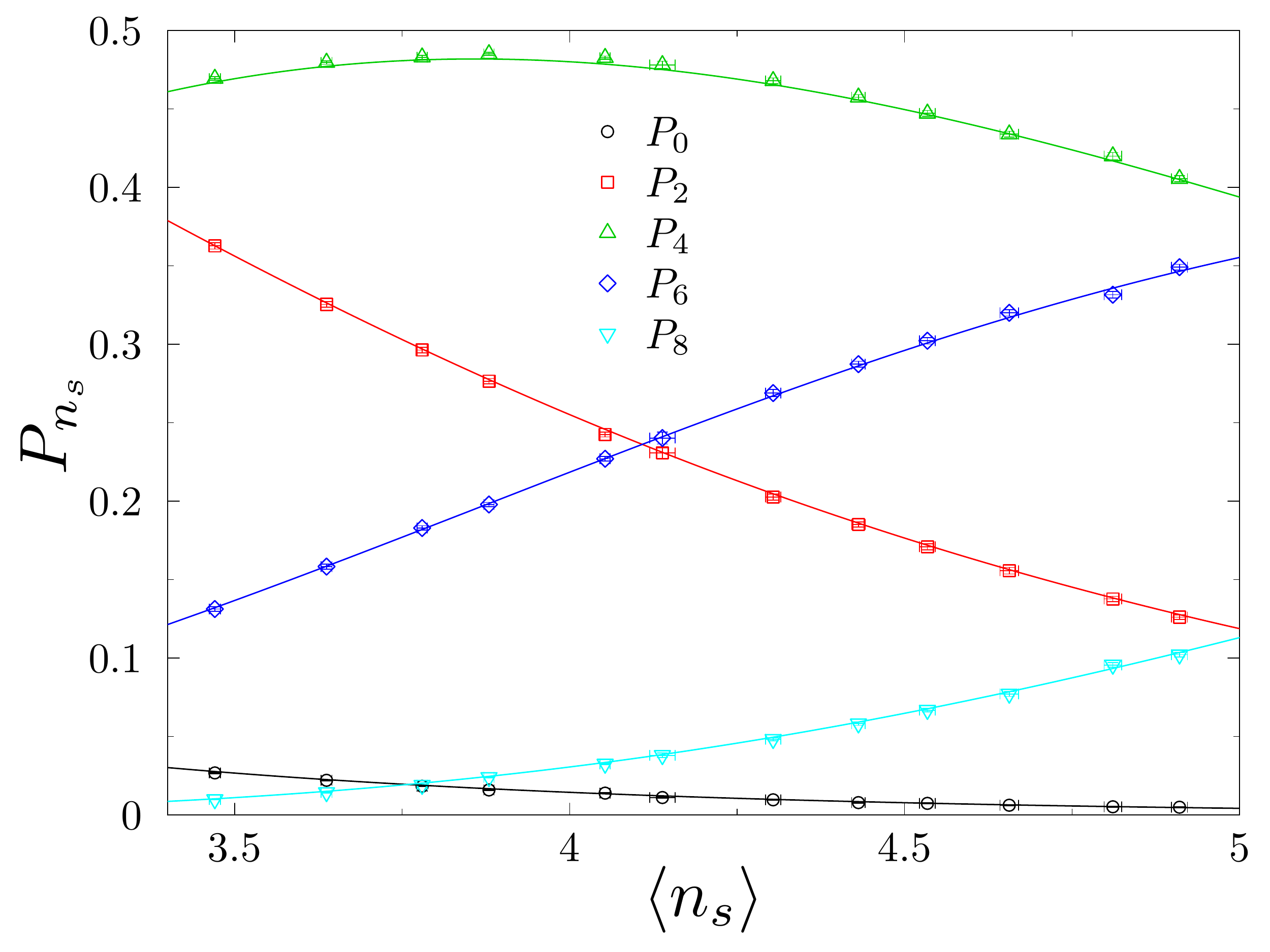}
\caption{Probabilities $P_{n_s}$ of $n_s$ spanning curves, as a function of $\<n_s\>$ (data for $p=q=1/2$). Curves are the analytical expressions from Eq.~\ref{Pns expression}, with $\widetilde K = \<n_s\>$.}
\label{Pns fig}
\end{figure}

Denoting the partition function with the above boundary conditions by $Z(\theta)$, we therefore have
\be
\< (\cos\theta)^{n_s} \> = \f{Z(\theta)}{Z(0)}.
\ee
In the Goldstone phase the stiffness flows to large values in the infra-red, so to calculate the right hand side we need consider only classical solutions with the appropriate boundary conditions. Letting $x$ be the coordinate along the cylinder, these are
\ba\notag
\vec S & = \pm (\cos \phi(x), \sin \phi(x), 0, \ldots, 0), &
\phi(x) & =\f{ x (\theta + \pi m)}{L}.
\end{align}
Both odd and even values of $m$ are allowed --- the boundary condition is satisfied only up to a sign ---  but when $L$ and $m$ are both odd the Boltzmann weight acquires an additional minus sign, as can be seen from the lattice partition function.\footnote{In the mapping of Sec.~\ref{Lattice field theory for the CPLC} this  sign comes from factors of $\vec S_\text{top}. \vec S_l$ and $\vec S_\text{bottom}. \vec S_l$ in the Boltzmann weight for links $l$ adjacent to the boundary links.}

The action of these solutions is calculated using the renormalized stiffness $\kt$ on scale $L$, 
leading to 
\ba
\label{generating function}
\< (\cos\theta)^{n_s} \> & \simeq \sum_m (-)^{m L} \exp \lf { - \f{\kt}{2} (\theta-\pi m)^2} \ri.
\end{align}
For a given value of $\theta$, only one or two values of $m$ are not exponentially small in $\kt$.

To extract low-order cumulants for $n_s$, we set $\cos \theta=e^{-x}$ and expand in $x$. Since  (\ref{generating function}) is dominated by the $m=0$ term for $\theta\sim0$, the difference between even and odd $L$ is not seen. The $l$th cumulant is given by
\be\notag
\<\< n_s^l \>\>  \simeq  - \f{ \kt }{2}  \partial_y^l   (\arccos e^{y})^2 \left.\right|_{y=0}.
\ee
In particular, the mean spanning number is given by the renormalized stiffness $\kt$, so
\be
\label{spanning number formula}
\< n_s\> \sim  \f{2-n}{2\pi} \ln  \f{L}{ L_0}.
\ee
This logarithmic flow (for $n=1$) is seen in Fig.~\ref{spanning number fig} for two points in the Goldstone phase. We have fitted the data for large sizes to the slightly more accurate form $\< n_s\> \simeq  \f{1}{2\pi}\lf \ln  {L}/{ L_0} +\ln  \ln  {L}/{ L_0} \ri$ which comes from including the subleading $O(1/K)$ term in the beta function for the stiffness (\ref{basic RG equation}). In the upper inset to Fig.~\ref{spanning number fig} we plot the numerical value of the slope $\dd \<n_s\>/\dd \ln L$, which is seen to converge slowly to $1/2\pi$ for large $L$.

Since all cumulants are proportional to $\kt$, their ratios are universal numbers which we can compare with data: 
\ba\notag
\<\< n_s^2 \>\>& = \f{2}{3} \<n_s\>, &\<\< n_s^3 \>\>& = \f{4}{15} \<n_s\>.
\end{align}
These relations are obeyed to good accuracy --- plotting the two cumulants above against $\<n_s\>$ for $p=1/2$ and various $L$ gives straight lines with slopes 0.668(5) and 0.274(18) respectively (data not shown). Note that the scaling of the cumulants means that when $\<n_s\>$ becomes very large the probability distribution $P_{n_s}$ for the spanning number becomes Gaussian (away from its tails).

To extract the probability distribution for small integer $n_s$ we set $\cos\theta=\epsilon$ in  (\ref{generating function}),
\be\label{Pns expression}
\sum_{n_s} P_{n_s} \epsilon^{n_s} \simeq
e^{- \f{\kt}{2} (\arccos \epsilon)^2 }  + (-)^L
e^{ - \f{\kt}{2} (\arccos \epsilon-\pi)^2 }.
\ee
For even circumference, this gives for example
\ba\notag
P_0 &= 2 e^{  - \pi^2 \<n_s\> /8 }, & \,\,\,
P_2 &=\f{\pi^2  \<n_s\>^2  - 4\<n_s\>}{4} e^{ - \pi^2  \<n_s\> /8}.
\end{align}
In Fig.~\ref{Pns fig} the expressions for $P_{0}, \ldots, P_8$ are compared with  data (at $p=q=1/2$ and $L$  in the range $10^2 - 10^6$) showing remarkable agreement. There is no free parameter in these fits.

\subsection{Length distribution}
\label{length distribution}

\begin{figure}[t] 
\centering
\includegraphics[width=3.3in]{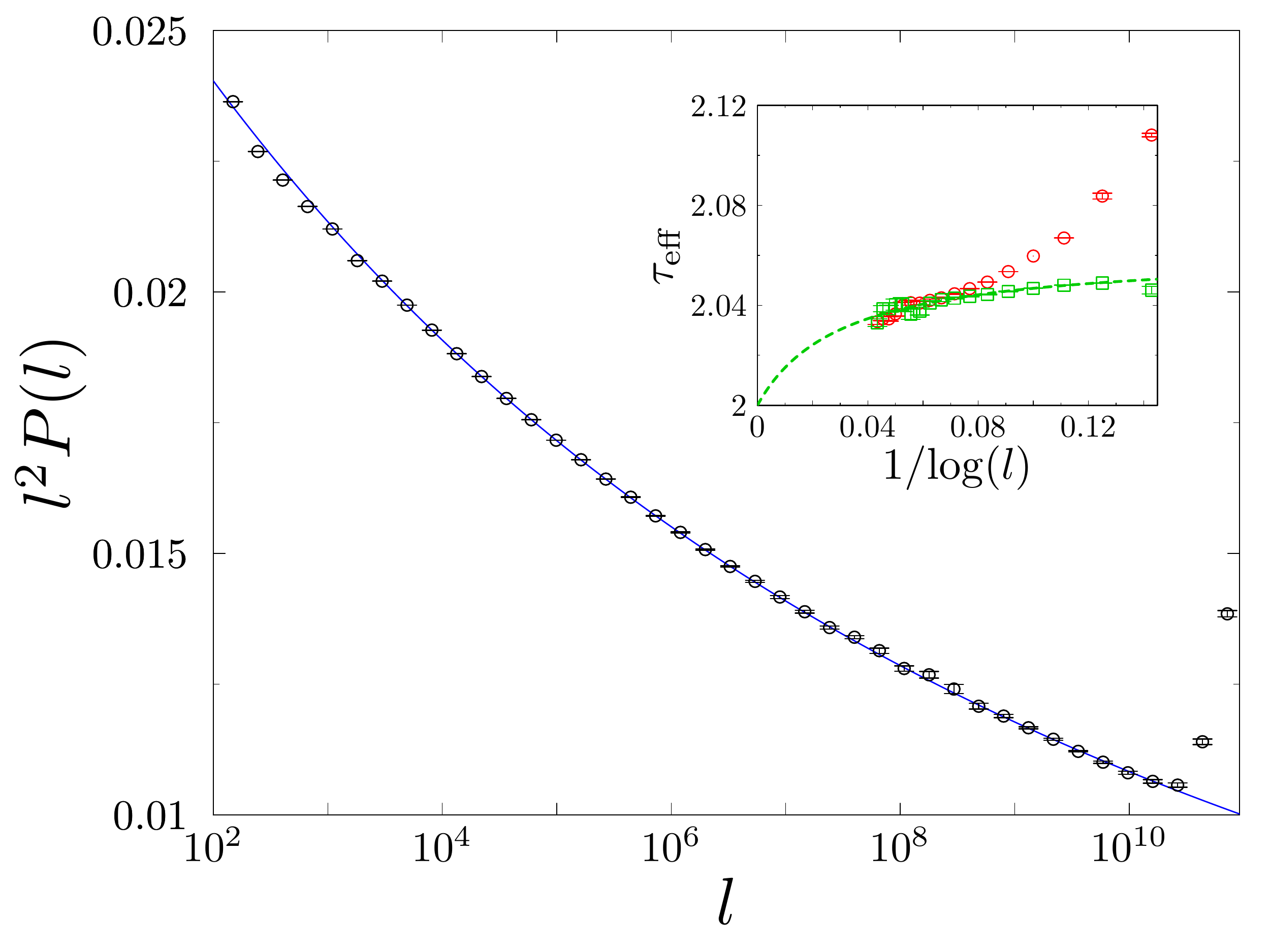}
\caption{Main panel: The probability distribution $P(l)$ for the length of a loop in the Goldstone phase. We have multiplied by $l^2$ to remove the expected power law part, leaving the logarithmic dependence. The fit is as described in the text, with $\ln l_0 = - 33.7(8)$. Data is for $p=q=1/2$. Inset: two ways of defining the effective finite-size exponent value $\tau_\text{eff}$ (see text). Green squares: data from $P(l)$, together with the fit (dashed line) implied by Eq.~\ref{P(l) formula} ($\ln l_0= - 32.7$). Red circles: data from $\Delta X(l)$.}
\label{extlengthdistribution}
\end{figure}

To calculate the length distribution for a loop we must consider the RG flow away from the Goldstone phase induced by a symmetry-breaking perturbation. To summarize the result of the following calculation, which is for $n=1$, the probability for a loop randomly chosen from the soup to have length $l$ falls off as
\be
\label{P(l) formula}
P(l) \propto \f{1}{l^2 \ln^2 (l/l_0)}
\ee
for large $l$.

Fig.~\ref{extlengthdistribution} shows the distribution obtained numerically for loops of length up to $l \sim 10^{10}$. We multiply $P(l)$ by $l^2$ in order to expose the logarithmic correction, which we fit to the form $a (\ln l / l_0)^{-c}$. We obtain 
\ba
c = 2.03(3)
\end{align}
in striking agreement with (\ref{P(l) formula}). This value is also in agreement with numerical results for trails \cite{Owczarek and Prellberg collapse, Ziff}, as we will discuss shortly.

Note that $P(l)$ differs by a factor of $l$ from the length distribution for the loop passing through a fixed link,
\be\label{P fixed link}
P_\text{fixed link}(l) \propto l \, P(l),
\ee
simply because longer loops visit more links. Thus  $\< l \>$ evaluated using $P(l)$ is finite, as for (\ref{P(l) formula}).

Let $g(x)$ denote the generating function for the length of a loop randomly chosen from the soup (the `marked' loop):
\be
g(x) = \sum_l P(l) x^l = \< x^\text{length of marked loop} \>.
\ee
In order to extract $g(x)$ we use the trick of Ref.~\cite{Cardy n+n' trick}, splitting the components of $\vec S$, or equivalently the loop colours, into two groups. For simplicity we will consider only the loop model at fugacity one, though it would be easy to generalize. We split $\vec S$ as in Eq.~\ref{S split}, yielding an $(n-1)$-component vector $\vec S_\perp$. In the graphical expansion of a lattice model, say of the CPLC (the IPLC would be similar) we correspondingly split the loops into unmarked loops, whose colour index is equal to one, and marked loops, whose colour index runs over $ 2, \ldots , n$. After summing over loop colours, a configuration with $N$ marked loops has a weight $(n-1)^N$, and expanding the partition function in $(n-1)$ is equivalent to expanding in the number of marked loops in the configuration. Writing $n'=n-1$,
\be  \label{n-1 expansion}
Z(n') = \sum_\mathcal{C} W_\mathcal{C} + 
n'
\hspace{-3mm}
 \sum_{\substack{\mathcal{C}; \text{ one} \\ \text{marked loop} }} 
 \hspace{-3mm}
 W_\mathcal{C} + \ldots
\ee
Here $W_\mathcal{C}$ is the weight of a configuration $\mathcal{C}$ in the CPLC at $n=1$ (Eq.~\ref{W weight}). The first term on the right hand side is equal to one.

Next, we wish to  modify the weight of a configuration by the factor $x^l$, where $l$ is the total length of marked loops. This is easily done: all the inner products $\vec S_l . \vec S_{l'}$ appearing in the node factors $e^{-E_i}$ (see Sec.~\ref{Lattice field theory for the CPLC}) are replaced according to:
\be
\vec S_l . \vec S_{l'} \quad \longrightarrow \quad 
S_{l }^1 S_{l'}^1 + x  \, \vec S_{l\perp}. \vec S_{l' \perp}.
\ee
In this way every unit length of marked loop picks up a factor of $x$, and the graphical expansion  yields
\be \label{n-1 expansion 2}
Z(n',x) = 1 + 
n'
\hspace{-3mm}
 \sum_{\substack{\mathcal{C}; \text{ one} \\ \text{marked loop} }} 
  \hspace{-3mm}
 W_\mathcal{C} \, x^{l} + \ldots
\ee
Differentiating with respect to $n'$ gives the required generating function:
\be
g(x) = \f{\partial_{n'} \left. Z(n', x) \right|_{n'=0} }{\partial_{n'} \left. Z(n', 1) \right|_{n'=0}}.
\ee
In terms of the free energy density $f(n',x)$,
\ba \notag
Z(n', x) & = e^{ - L^2 f(n',x)}, &
f(n',x) & = f_0 + n' f_1(x) + \ldots,
\end{align}
this is $g(x)  = {f_1(x)}/{f_1(1)}$. In order to calculate the length distribution for large values of $l$, we require the free energy for $x \simeq 1$ and to first  order in $n'$.

In the continuum description, the symmetry-breaking perturbation $x \neq 1$ leads to an infinite number of relevant perturbations of which the most relevant is that in Eq.~\ref{perturbed lagrangian}. Setting $x= \exp ( - \mu)$ with $\mu \ll 1$, we have $\gamma \sim \mu$, the constant of proportionality being nonuniversal.

Beginning with the Lagrangian of Eq.~\ref{perturbed lagrangian}, we integrate out high frequency modes, retaining their contribution to the free energy, up to an RG time $\tau_*$. This gives
\ba\label{f split}
f(K, \gamma) =  f^{<}(K,\gamma)+  f^{>}(K,\gamma),
\end{align}
where we have split up the contribution from the modes that have been integrated out,
\ba \notag
f^{<}(K,\gamma)  =  \f{n'}{4\pi}
\int_0^{\tau_*} \f{\dd \tau}{e^{2\tau}} \lf \ln K(\tau) + \gamma(\tau)  \ri  - n' A
\end{align}
(the nonuniversal constant $A$ ensures $f<0$, as required by Eq.~\ref{n-1 expansion 2}) and those remaining:
\ba
 \notag
f^{>}(K,\gamma) & = e^{-2\tau_*} f(K(\tau_*), \gamma(\tau_*)).
\end{align}
The solutions to the RG equations (\ref{K gamma RG equations}) are
\ba
\label{solution to RG equations}
K(\tau) &= K + \f{\tau}{2\pi}, &
\gamma(\tau) &= \gamma e^{2\tau} \lf \f{2\pi K}{2\pi K + \tau} \ri^2.
\end{align}
Stopping the RG when $\gamma_*$ becomes of order one,  $f^>$ may be approximated as the free energy of a massive theory in which $\vec S$ executes only small fluctuations around $\vec S = (1,0,\ldots, 0)$: 
\be
\label{massive renormalized lagrangian}
\mathcal{L} = \f{K_*}{2}\lf (\nabla \vec S_\perp)^2 +  \gamma_* \vec S_\perp^2 + O(\vec S_\perp^4, \, \vec S_\perp^2/K_*) \ri.
\ee
(The $O(\vec S_\perp^2/K_*)$ term comes from the sigma model measure.) However, the dominant terms in $f$ come from $f^<$. To the order that we require,
\be
f(K, \gamma) = -B + \gamma \lf   2\pi K - \f{8 \pi^2 K^2}{\ln 1/\gamma} + \ldots \ri
\ee
($B$ is a constant.) We thus have the form of the generating function at small $\mu$:
\be
\< e^{-\mu l} \> = 1 -  C \, \mu \lf 1 - \f{4\pi K}{ \ln 1 / \mu}  + \ldots \ri.
\ee
The constants $C$ and $K$ are non-universal, and the fact that the leading $\mu$ dependence is linear in $\mu$ is simply a consequence of the fact that $\<l \>$ is finite. However we may infer the behaviour of $P(l)$ at large $l$ from the form of the nonanalytic term, yielding Eq.~\ref{P(l) formula}.

Previous work on self-avoiding trails \cite{Owczarek and Prellberg collapse, Ziff}, which map to the $n=1$ loop model at $p=1/3$, $q=1/2$ (see Sec.~\ref{polymer section}), considered a probability $Q(l)$ which may be written
\be \label{Q(l)}
Q(l) = \int_l^\infty P_\text{fixed link} (l') \dd l'.
\ee
Viewing the loops as the trajectories of walkers, $Q(l)$ is the probability that a walker has not yet returned to its starting point after $l$ steps. Eqs.~\ref{P(l) formula},~\ref{P fixed link},~\ref{Q(l)}  give $Q(l) \sim 1/ \ln l$. This agrees with the scaling found numerically in Refs.~\cite{Owczarek and Prellberg collapse, Ziff}.

For a generic critical loop ensemble, $P(l)\sim l^{-\tau}$ for some $\tau\geq 2$, and the mean size $\Delta X$ of a loop scales with its length as $\Delta X\sim l^{1/d_f}$. The fractal dimension $d_f$ is related to $\tau$ by 
\be
\label{tau scaling relation}
\tau = 2/d_f+1.
\ee
In the Goldstone phase, $\tau=d_\text{eff}=2$, with logarithmic corrections. We may define  finite size estimates of $\tau$  either using  $\dd \ln P / \dd \ln l$ or using  $\dd \ln \Delta X/\dd \ln l$ and the scaling relation. These are plotted in the inset to Fig.~\ref{extlengthdistribution}; both are expected to converge to two, but with different logarithmic corrections. 

Here, $\Delta X$ is defined as the mean extent of a loop in one of the coordinate directions. A similar quantity --- the mean square end to end distance of an open trail in the ISAT model --- was considered numerically by Owczarek and Prellberg \cite{Owczarek and Prellberg collapse}, and logarithmic corrections to Brownian scaling were found. It would be interesting to calculate these quantities analytically.

\section{The critical lines}
\label{critical line}

The critical lines separate the Goldstone phase from phases with short loops. In the language of the $\rp^{n-1}$ model, they correspond to order-disorder transitions at which $\mathbb{Z}_2$ vortices are set free. In this section we give numerically determined critical exponents for this transition at $n=1$ and briefly consider an approximate RG treatment of vortices \cite{Fu Kane, Konig et al}.

\subsection{Critical spanning number, $\nu$, and $y_\text{irr}$}
\label{critical spanning number}

\begin{figure}[t] 
\centering
\includegraphics[width=3.35in]{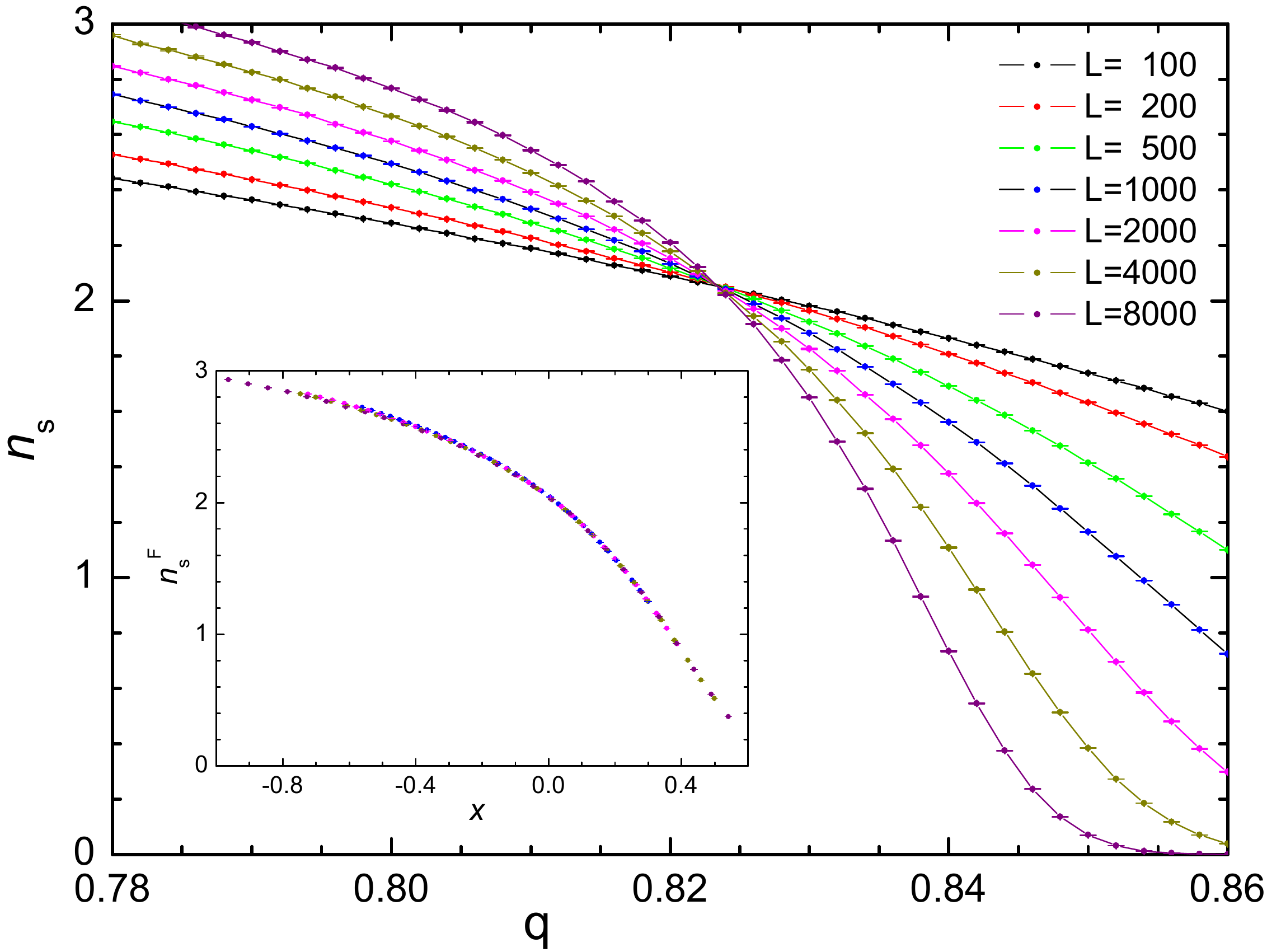}
\caption{Main panel: the mean spanning number $n_s$ as a function of $q$ for $p=1/2$ and various system sizes, showing a crossing at the transition. Inset: data collapsed according to Eqs.~\ref{scaling variable},~\ref{finite size effects}.}
\label{crossingsfig1}
\end{figure}

\begin{figure}[t]
\centering
\includegraphics[width=3.35in]{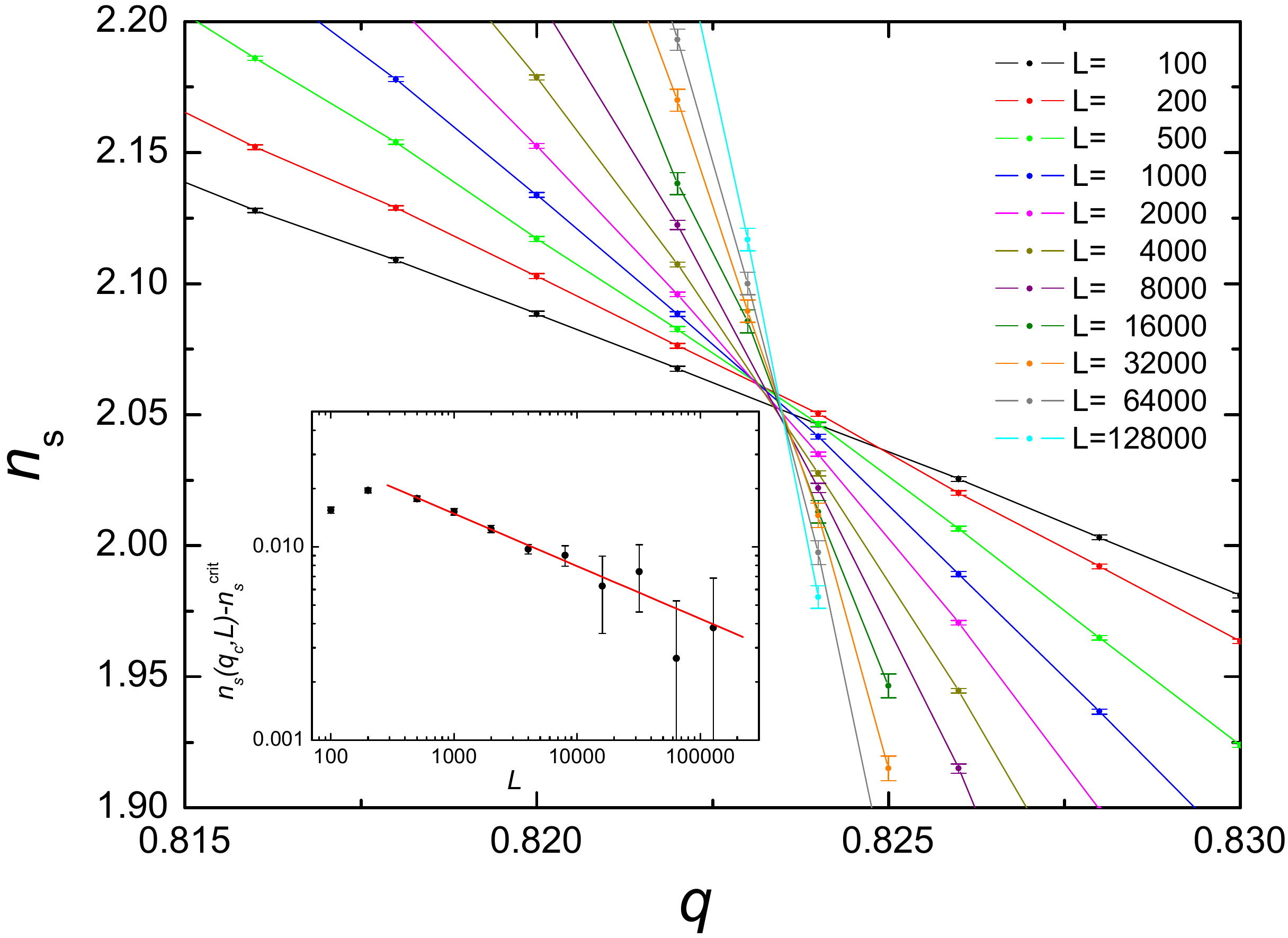}
\caption{Main panel:  the behaviour of the mean spanning number at $p=1/2$  very close to the critical point. Note the larger system sizes compared to  Fig.~\ref{crossingsfig1}. Inset: finite size corrections to the spanning number at the critical point (note log-log scale) and linear fit leading to estimate of $y_\text{irr}$.}
\label{crossingsfig2}
\end{figure}

\begin{figure}[t]
\centering
\includegraphics[width=3.35in]{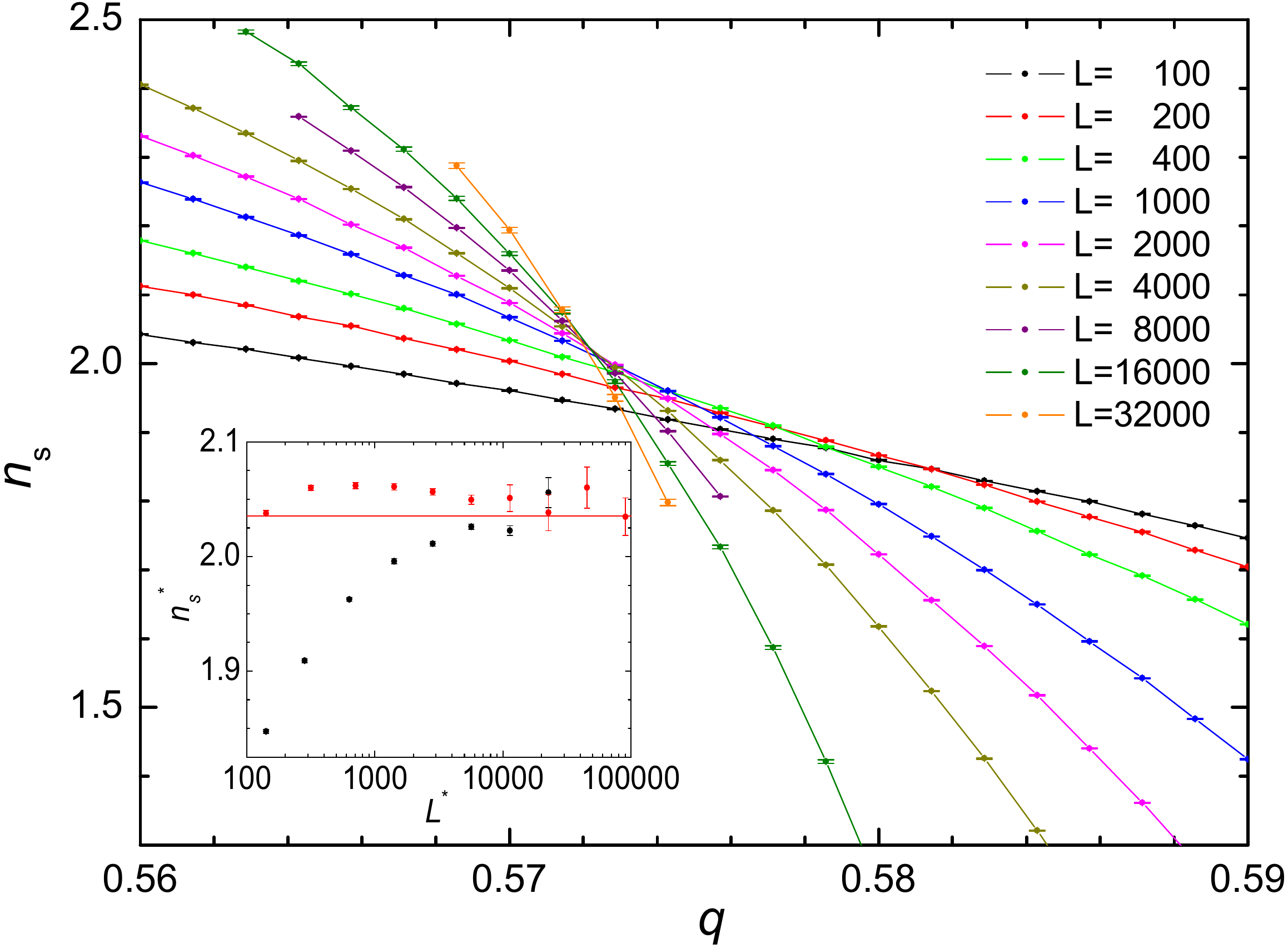}
\caption{Main panel:  spanning number at $p=0.3$ (close-up of the critical point) showing larger drift in crossing points than at $p=1/2$ (however, note different scale). Inset: the vertical coordinate $n_s^*$ is the crossing in the spanning number between consecutive system sizes and $L^*$ is the geometrical mean of the sizes. Red and black dots correspond to $p=1/2$ and $p=0.3$ respectively. The horizontal line is the estimated asymptotic value (\ref{nu and nscrit}).}
\label{crossingsfig3}
\end{figure}

 At the critical point, the dimensionless quantity $n_s$ (defined in Sec.~\ref{spanning number section}) is expected to take a universal value. This is manifested in the  crossings of the various curves in Fig.~\ref{crossingsfig1}, which shows the spanning number $n_s$ as a function of $q$ for $p=1/2$ and for cylinders of various sizes. Fig.~\ref{crossingsfig2} shows the same quantity, but in the immediate vicinity of the critical point and including much larger system sizes (up to $L=128,000$). The main panel of Fig.~\ref{crossingsfig3} shows data for $p=0.3$; here finite size effects --- visible in the drift in crossings --- are much stronger.

The  basic finite size scaling form for $n_s$ is
\ba
n_s &= h(x), & x& = L^{1/\nu} \delta q,
\end{align}
where $\delta q = q - q_c$. We take into account also nonlinear dependence of the scaling variable $x$ on $\delta q$, replacing the second equation above with
\ba\label{scaling variable}
x =  L^{1/\nu} \delta q \lf 1 + \beta_1 \delta q + \beta_2 \delta q^2 \ri,
\end{align}
and finite size corrections with (negative) irrelevant exponent $y_\text{irr}$ in the form:
\ba\label{finite size effects}
n_s = h(x) \lf 1 + L^{y_\text{irr}} (\beta_3 + \beta_4 x)  \ri.
\end{align}
A reasonable scaling collapse may be obtained by adjusting the values of $q_c$, $\beta_i$, $\nu$ and $y_\text{irr}$. To find these values we fit $n_s$ to the form (\ref{finite size effects}), constructing  $h(x)$ using B-splines with 22 degrees of freedom. The result for $p=1/2$ is shown in the inset to Fig.~\ref{crossingsfig1}. What is plotted is  $n_s^F = n_s / \lf 1 + L^{y_\text{irr}} (\beta_3 + \beta_4 x)  \ri$, which should be equal to the scaling function $h(x)$ by (\ref{finite size effects}). 

Our estimates of the correlation length exponent and (universal) critical spanning number, obtained from the data at $p=1/2$, are:
\ba \label{nu and nscrit}
\nu & = 2.745(19) & 
n_s^\text{crit} & = 2.035(10).
\end{align}
We cannot constrain the irrelevant exponent very precisely. From the full fit, we obtain
\be
\label{irrelevant exponent}
y_\text{irr} \in - (0.2 , 0.35).
\ee
A direct estimate from the finite size corrections to the spanning number at the critical point gives a result compatible with this: the fit in the inset to Fig.~\ref{crossingsfig2} corresponds to $y_\text{irr}=-0.272$.

Results for $p=0.3$ are consistent with our expectation that all points on the critical line are in the same universality class, but error bars are larger because of larger finite size effects and smaller system sizes. We find $\nu = 2.87(10)$ and $n_s^\text{crit} = 2.07(3)$.  With regard to the convergence to a common $n_s^\text{crit}$, see the inset to Fig.~\ref{crossingsfig3} which shows the vertical coordinates of the crossings between curves for consecutive $L$ values.

\subsection{Watermelon correlators,  fractal dimension, and length distribution}

\begin{figure}[t] 
\centering
\includegraphics[width=3.4in]{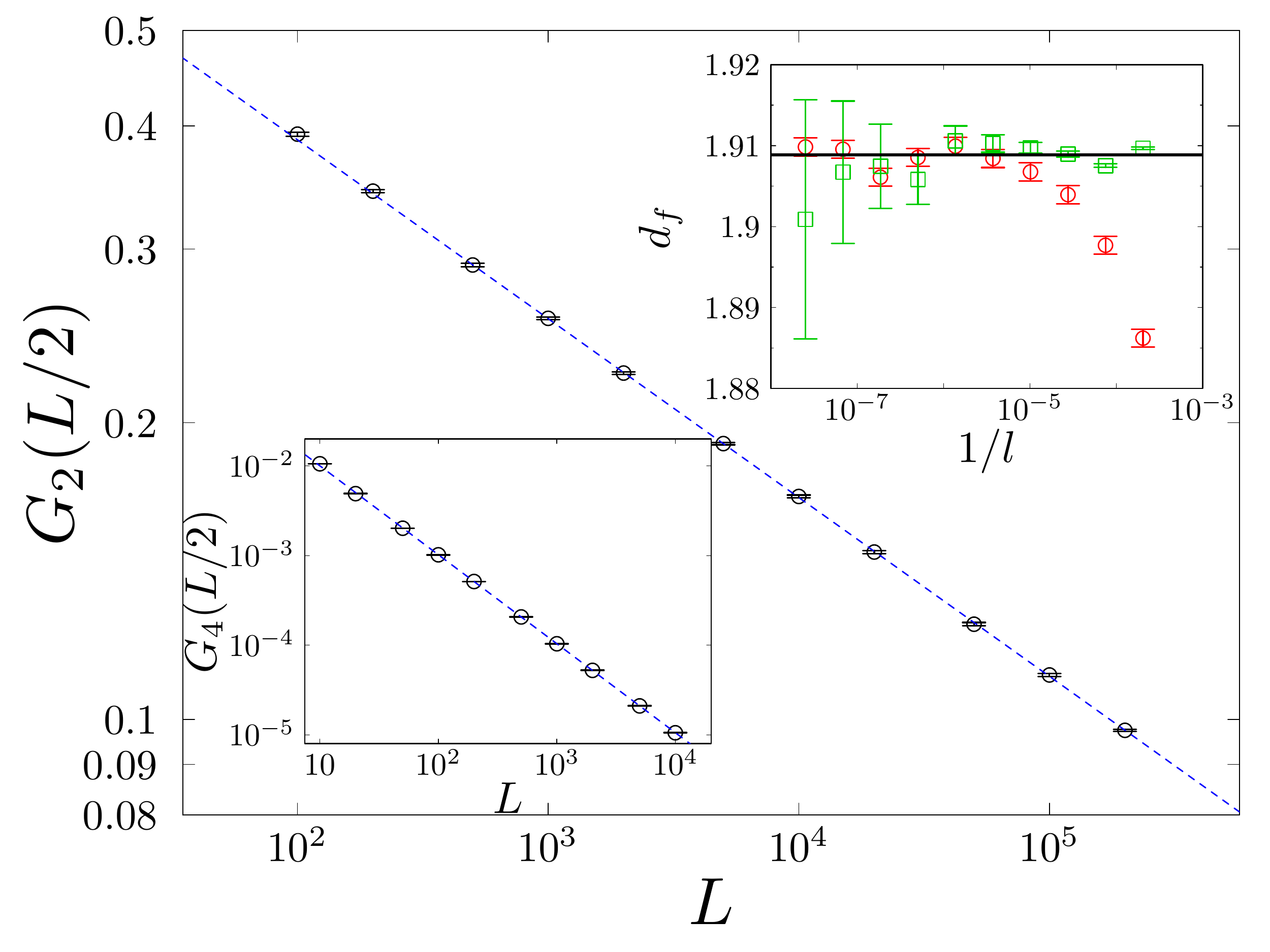}
\caption{Main panel: the two-leg watermelon correlator $G_2$ at the critical point. The power-law decay $G_2\sim L^{-2x_2}$ gives the fractal dimension $d_f$ via (\ref{df scaling relation}). The upper inset compares this value (indicated by the horizontal black line) with the finite-size estimates coming from $\Delta X(l)$ (red circles) and $P(l)$ (green squares) --- see text. The lower inset shows the four-leg watermelon correlator $G_4$.}
\label{G2critical}
\end{figure}

Next we consider the watermelon correlation functions $G_2$ and $G_4$ defined in Sec.~\ref{correlation functions and replica}. We evaluate these correlators at separation $L/2$ for a range of system sizes $L$ -- see Fig.~\ref{G2critical}. The data is for the critical point at $p=1/2$.

Fitting to pure power laws, 
\be
G_k(L/2) \propto L^{- 2 \, x_k},
\ee
we obtain the scaling dimensions of the two- and four-leg operators:
\ba
x_2 & = 0.091(1), & x_4 & = 0.491(1).
\end{align}
The scaling relation
\be
\label{df scaling relation}
d_f = 2 - x_2
\ee
gives the fractal dimension of the critical loops, 
\be
\label{df from x2}
d_f = 1.909(1).
\ee
We may obtain independent estimates of  $d_f$ from $\Delta X (l)$, the mean linear size of a loop of length $l$, and from the length distribution $P(l)$ and the scaling relation (\ref{tau scaling relation}). The finite size estimates for $d_f$ coming from the numerical estimates of  $\dd \ln \Delta X / \dd \ln l$ and $\dd \ln P(l) / \dd \ln l$ are shown in the inset to Fig.~\ref{G2critical}. Both plots are consistent with (\ref{df from x2}).

\subsection{RG equations in the presence of vortices}
\label{RG equations with vortices}

For an approximate description of the transition, we  extend the RG description (\ref{basic RG equation}) to take account of the nonzero fugacity for $\mathbb{Z}_2$ vortices. In this we follow the treatment by Fu and Kane of the $O(2N)/O(N)\times O(N)$ sigma model at $N\rightarrow 0$ \cite{Fu Kane}. This sigma model and the $\rp^{n-1}$ sigma model are similar --- both sustain $\mathbb{Z}_2$ vortices, and each reduces to the XY model in an appropriate limit, which can be expanded around. An expansion around the XY limit was also considered for the $O(n)$ model near $n=2$ in Ref. \cite{Cardy Hamber}. The importance of topological defects in replica sigma models for localisation in two dimensions was also pointed out by Konig et al., who developed an RG approach to localisation in the chiral symmetry classes taking account of $\mathbb{Z}$ vortices \cite{Konig et al}.

Since $\rp^1=S^1$, our sigma model coincides with the XY model at $n=2$. With the normalisation of Eq.~\ref{introduction of sigma model}, this has a Kosterlitz Thouless transition at the critical stiffness $K_c=8/\pi$. The RG flow near this point is governed by the Kosterlitz RG equations for $K$ and the vortex fugacity, which we denote $V$. 

We assume that we can expand the  RG equations in $(2-n)$, and that $V$ should be interpreted as the fugacity for $\mathbb{Z}_2$ vortices\footnote{For $n>2$, a $\mathbb{Z}_2$ vortex corresponds to e.g. $\vec S = (\cos\theta/2, \sin\theta/2, 0, ..., 0)$; note that when $n=2$ this configration becomes a $\mathbb{Z}$ vortex of charge $\pm 1$. (Here $\theta$ is a polar coordinate; recall $\vec S\sim -\vec S$.)} when $n \neq 2$. At lowest order, these equations are corrected by the $\beta$-function for $K$ in the absence of vortices, $\dd K/\dd \tau \simeq (2-n) f(K)$:
\ba \label{mu flow}
\f{\dd V}{\dd \tau} & = \lf 2  - \f{\pi K}{4} \ri V,  \\ \label{K flow}
\f{\dd K}{\dd \tau} & = (2-n) f(K_c) - V^2.
\end{align}
These equations yield critical points at  $K=K_c=8/\pi$ and $V = \pm \sqrt{(2-n)f(K_c)}$, with critical exponents
\ba\notag
\nu &= \sqrt{\f{2}{(2-n)\pi f(K_c)}}, &
y_\text{irr} & = - \sqrt{\f{(2-n)\pi f(K_c)}{2}}.
\end{align}
The critical stiffness $K_c\sim 2.4$ is of roughly the same magnitude as the critical winding number $n_s^\text{crit}\sim 2.0$ at $n=1$ (\ref{nu and nscrit}), as we expect from Sec.~\ref{spanning number section}. Making the further approximation of evaluating $f(K_c)$ using the perturbative $\beta$-function at large $K$,
\be\notag
f(K) \simeq \f{1}{2\pi} \lf 1 + \f{1}{2\pi K } \ri,
\ee
yields at $n=1$
\ba\notag
\nu & \sim 1.9, &y_\text{irr} \sim -0.5.
\end{align}
As $n\rightarrow 2$, $\nu$ diverges and the irrelevant exponent tends to zero.

As expected, this crude approximation does not give quantitatively accurate results for $n=1$, but it does reproduce the qualitative structure of the phase diagram, with the Goldstone phase sandwiched between massive phases at positive and negative $V$, and the appearance of a large correlation length exponent. 

A comparison with alternative approaches --- for example an approximate treatment of sigma model directly at $n=1$ in the supersymmetric formulation, or an expansion in $(2-n)$ avoiding the additional large $K$ approximation required here --- would be desirable.

\section{Numerical methods}
\label{Numerical methods}

We have considered system sizes from $L=100$ up to $L=10^6$. For sizes up to $L \sim 2\times 10^4$ we can use a straightforward Monte Carlo procedure, which of course benefits from the fact that node configurations are independent random variables when $n=1$. Very large sizes require a more efficient `knitting' procedure.

In the straightforward approach, we construct independent $L\times L$ samples, assigning the node configurations at random with the probabilities in Sec.~\ref{CPLC}. By following the loops we then calculate the spanning number, $n_{\rm s}$; the loop length distribution, $P(l)$; the average linear size of a loop of length $l$,  $\Delta X(l)$; and the correlation functions $G_2$ and $G_4$ at separation $L/2$.   The spanning number (defined in Sec.~\ref{spanning number section}) requires open cylinder boundary conditions, while fully periodic BCs are used for the other observables. However the same samples may be used for both, since the Boltzmann weight is independent of the boundary conditions when $n=1$. 

The two-leg correlator $G_2(L/2)$ is the probability that two links at separation $L/2$ lie on the same loop, and we take $G_4(L/2)$ to be the probability that two nodes at separation $L/2$ are visited by the same pair of loops. This is not the only way that two nodes can be joined by four strands --- the four strands can make up a single loop rather than a pair --- but the scaling is the same for the two types of contribution.

The data for $P(l)$ and $\Delta X(l)$ are stored in histograms in a logarithmic scale in $l$. For each box we calculate the mean and standard deviation of the loop size. The former gives the estimate of $\Delta X(l)$; to avoid finite size effects, a box is discarded if its $\Delta X$ is within two standard deviations of $L$.

With this procedure, the number of independent samples constructed for a given system size varied between $2\times 10^5$ for the largest size and $2\times 10^6$ for the smallest.

\subsection{Knitting}

As we expect  logarithmic behaviour in the Goldstone phase, it is crucial to be able to study very large systems. The straightforward procedure above is very efficient in terms of CPU time, but it is limited by the available computer memory since it requires us to store  the full configuration. The first improvement comes from the fact that a cylinder of circumference $L$ can be `knitted' by the successive addition of $L \times L'$ strips, where $L'$ is of order one.

At a given stage in the growth of the cylinder, it contains both closed loops in the interior, and strands which end on the boundary. The lengths and sizes of the closed loops are already included in the histograms for $P(l)$ and $\Delta X(l)$. Then, the only information about the configuration which we need to store is the connectivity of the links on the boundary, together with the lengths of the connecting strands. (Note that this connectivity information is configuration-dependent --- this is distinct from the transfer matrix approach, in which the transfer matrix is not configuration-dependent, and which is limited to small sizes.) When we add a new strip to the cylinder some loops will become closed: we add their lengths $l$ and their sizes $\Delta X$ -- defined as the height of the loop in the growing direction -- to our histograms. 

Growing the cylinder strip by strip, the required memory is proportional to the circumference $L$ (strictly to  $L\ln L$) rather than to $L^2$ as in the straightforward approach (algorithms for percolation  which avoid storing the full configuration also exist  \cite{hoshen kopelman}). We have been able to compute cylinders with $L$ up to $10^6$ and height much greater than $L$.  

This method requires storing the (realisation-dependent) connectivity information for the boundary links of a large cylinder. This has the flavour of a transmission matrix in a localisation problem. In the future it would be interesting to consider the properties of this `matrix' in more detail.

\subsection{Shuffling}

A further improvement that significantly reduces CPU effort involves constructing sets of strips of width $L$ and height $H=L/20$ (using the knitting procedure). For each strip we store the connectivity of the boundary links. Joining 20 of them yields a square sample, and we have enough information about this sample to calculate $G_2(x,y)$ (for links $x$ and $y$ which lie on the boundaries of two strips) as well as $n_s$. From each set of 20 pieces, many different samples may be created by shuffling the order of the pieces and by rotating them in the  transverse direction.  We construct 1000 samples for each set. These samples are not of course independent, so we estimate error bars by producing many independent such sets ($80-200$) and examining the statistical fluctuations between sets.
  
For the calculation of $n_s$ in large systems to a given precision, the shuffling and rotation of strips reduces CPU time by a factor of 200.

We have parallelized this procedure for OpenMP and for CUDA GPUs. The parallelization for the graphics cards was particularly efficient -- a typical programme ran almost 100 times faster on an Nvidia Tesla M2070 card than on a single core in an Intel Xeon E5520 CPU.

\section{Polymer collapse}
\label{polymer section}

\begin{figure}[b] 
\centering
\includegraphics[width=3.5in]{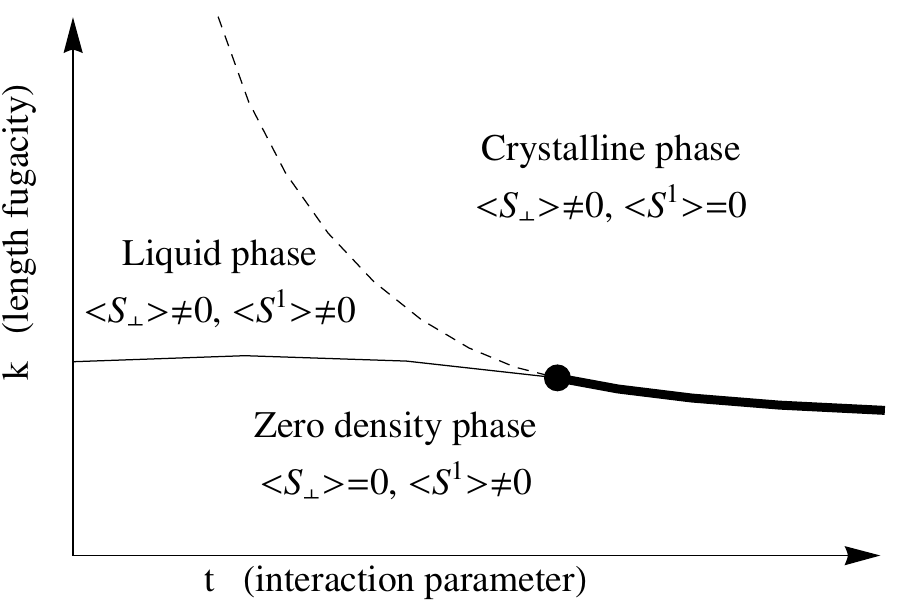} 
\caption{Schematic representation of the phase diagram for the polymer found in Ref.~\cite{foster universality}, together with our interpretation in terms of a perturbed $O(1+n')$ model in the limit $n'\rightarrow 0$. The thin solid line is in the universality class of the self-avoiding walk (polymer in good solvent). The dot is the multicritical collapse point, or $\Theta$ point, with full $O(1+n')$ symmetry.  The bold line is a first-order transition. The dashed line is a transition in the Ising universality class between two regimes in which the polymer is dense (visits a finite fraction of the links of the lattice) -- the `crystalline' and `liquid' phases.}
\label{polymer phase diagram}
\end{figure}

A long polymer with repulsive (or excluded volume) interactions between segments displays the universal behaviour  of the lattice self-avoiding walk (SAW), while  strong enough attractive interactions cause the polymer to collapse. The boundary between these two regimes is the so-called $\Theta$ point. In de Gennes' description of the polymer via the $O(N\rightarrow 0)$ model, the SAW corresponds to the critical point, and the $\Theta$ point to the tricritical point \cite{replica for loops, de gennes tricritical, cardy book}. However the actual situation is more complicated, especially in two dimensions.

While the SAW behaviour is extremely robust, the $\Theta$ point is more subtle. Different lattice models can yield different universality classes of collapse transition  \cite{Blote Nienhuis, duplantier  saleur polymer, Owczarek and Prellberg collapse, foster universality}, and it also turns out that the phase diagram in the vicinity of the $\Theta$ point can have a more complex structure than would naively have been expected \cite{Blote Nienhuis, foster universality, Doukas triangular lattice, guo blote nienhuis}. 

The interacting self-avoiding trail model (ISAT) has been particularly controversial \cite{Lyklema, Guha, Meirovitch, Owczarek and Prellberg collapse, foster universality, Shapir Oono}, with numerous conflicting results and hypotheses put forward for the critical exponents at the collapse point. In this model, the polymer can visit links only once, but nodes twice, so allowed configurations of a closed polymer are  equivalent to allowed configurations of a single loop in the CPLC. Here we give a field theoretic description of the ISAT which explains the phase diagram found numerically \cite{foster universality} and shows that the ISAT $\Theta$ point is highly fine-tuned from the point of view of more general polymer models, being an infinite-order multicritical point at which the $O(N)$ symmetry of the problem is enhanced from the generic $O(N\rightarrow 0)$ to $O(N\rightarrow 1)$.

On a lattice of large finite size $L$, the ISAT partition function is\footnote{Translated versions of a loop are counted as distinct polymer configurations in the above sum.}
\ba\label{Zpol}
Z_\text{pol} (k, t)
=  \sum_{\substack{\text{polymer}\\\text{configs}}}
k^\text{length}
t^\text{no. self-contacts}.
\end{align}
A self-contact is a node visited twice by the polymer. We are now at fixed length fugacity rather than fixed length, but the two ensembles are simply related.\footnote{Fixed length fugacity $k$ is more convenient from the point of view of field theory and also if we wish to consider phases in which the polymer becomes space-filling. The two ensembles are related by a Laplace transform. The critical properties of a polymer of fixed large length on an infinite lattice are governed by the singularity of the partition function closest to $k=0$ \cite{cardy book} --- i.e. in the case of Fig.~\ref{polymer phase diagram} by a point on the line separating the zero-density phase from the dense phase. The `crystalline' and `liquid'  phases are dense, i.e. the length of the polymer is comparable with the lattice area.} The parameter $t$ controls interactions which are repulsive for $t<1$ and attractive for $t>1$.

An interesting feature of this model is the phase diagram,  obtained numerically by Foster \cite{foster universality}. A schematic version is shown in Fig.~\ref{polymer phase diagram}. For small $k$ the polymer is of finite typical size and not critical (the `zero density' phase), while for large $k$ it is dense, i.e. has length of order $L^2$. When $t$ is small the transition between these phases shows the usual critical behaviour of the SAW (thin solid line), and for large $t$ there is a first order transition associated with the collapsed polymer (thick line). The $\Theta$ point separates these. 

An unexpected feature, from the point of view of the de Gennes theory, is an additional line of transitions within the dense phase (dashed line) which are found numerically to be in the Ising universality class \cite{foster universality}. A similar line of Ising transitions terminating at a $\Theta$ point was found in a polymer model without crossings studied by Bl\"ote and Nienhuis \cite{Blote Nienhuis}, and a heuristic explanation was provided by associating Ising degrees of freedom with the faces of the lattice \cite{Blote Nienhuis, Nienhuis communication}, for which the polymer was a domain wall. In that model, the absence of crossings also allows for a Coulomb gas description which captures the Ising transition \cite{Jacobsen Kondev Transition}.

For a field-theoretic description of the ISAT, we make use of the fact that precisely at the $\Theta$ point, which is at $k=1/3$, $t=3$ \cite{Owczarek and Prellberg collapse}, the ISAT maps to the CPLC at a point in the Goldstone phase, namely $n=1$, $p=1/3$, $q=1/2$.  The $\Theta$ point is therefore described by the $O(n\rightarrow1)$ sigma model \cite{dense loops and supersymmetry}, and we can understand the region around it by perturbing this sigma model. The following considerations may readily be generalised to the various modifications of ISAT that have been studied numerically, e.g. on other lattices \cite{Doukas triangular lattice}, with additional interactions \cite{ISAT extra int, foster generalized, foster surface behaviour}, or in higher dimensions \cite{Prellberg and Owczarek 3D}.

Before continuing, we note that an alternative conjecture for the critical behaviour of the ISAT $\Theta$ point was put forward on the basis of numerical transfer matrix calculations in Ref.~\cite{foster universality}. According to this conjecture, the $\Theta$ point has nontrivial critical exponents identical to those of an exactly solvable model of polymers \emph{without} intersections \cite{Blote Nienhuis, Warnaar Batchelor Nienhuis}. This is at odds with the predictions of the Goldstone phase, which yields trivial critical exponents together with universal logarithmic corrections. We believe the Goldstone phase scenario is convincingly established by our numerical results for large systems (note that Sec.~\ref{spanning number section} includes data at the relevant point $p=1/3$, $q=1/2$), together with the logarithmic behaviour seen numerically in  Refs.~\cite{Owczarek and Prellberg collapse, Ziff} and the theoretical arguments of Ref.~\cite{dense loops and supersymmetry} and Sec.~\ref{lattice field theory section}, and that the apparent nontrivial exponents in Ref.~\cite{foster universality} are due to logarithmic finite size corrections (see endnote for more details\footnote{We can use the sigma model to estimate the eigenvalues of the transfer matrix for a cylinder of circumference $L$. Coarse-graining by a factor $L$ yields an effective 1D sigma model with renormalised stiffness $\tilde{K}_L$. This is equivalent to  the quantum mechanics of a particle on an $(n-1)$-dimensional sphere, with energy levels $E_k  = k(k-1)/ (2\tilde K_L)$, $k = 0,1,2,\ldots$ in the limit $n\rightarrow 1$. This leads to the finite size estimates $x_k = E_k / 2\pi$ for the scaling dimensions, which tend to zero logarithmically with $L$. For a very crude estimate at small sizes, we can approximate $\tilde K_L$ by the spanning number on an $L\times L$ cylinder, which we have calculated numerically at $p=1/3$. For $L$ from $5$ to $12$, this gives $x_2$ varying from $0.103$ to $0.086$.}).

To describe the single-polymer problem in the language of the spin model, we proceed along similar lines to Sec.~\ref{length distribution}, expanding the partition function for the CPLC in $n'=n-1$ to separate out configurations with a single marked loop (which will be our polymer).

To control $k$ and $t$ for this polymer we must modify the partition function (\ref{ZforCPLC}) in a way that breaks the symmetry of the $O(n)$ sigma model down to $\mathbb{Z}_2\times O(n')$. At each node, (\ref{ZforCPLC})  contains  terms of the form $(\vec{S}_1. \vec{S}_2) (\vec{S}_3. \vec{S}_4)$. Such a term corresponds to two sections of loop passing through the node, one connecting link $1$ to link $2$, and one connecting link $3$ to link $4$. In the new ensemble, these sections can be sections of marked or unmarked loop, and we modify the weights accordingly:
\ba \label{polymer node}
(\vec{S}_1. \vec{S}_2) &(\vec{S}_3. \vec{S}_4) \longrightarrow  \\ \notag
& \phantom{+} (S^{1}_1 S^{1}_2) (S^{1}_3. S^{1}_4)\\ \notag & +
3k (S^{1}_1. S^{1}_2) (\vec S_{\perp3}. \vec S_{\perp4})+
3k (\vec S_{\perp1}. \vec S_{\perp2}) (S^{1}_3. S^{1}_4) \\ \notag
&+  3 k^2 t\, (\vec S_{\perp1}. \vec S_{\perp2}) (\vec S_{\perp3}. \vec S_{\perp4}) .
\end{align}
Each unit of marked length acquires a factor of $3k$, and each meeting of two marked strands acquires an additional factor of $t/3$. Expanding in $n'$ as in Eqs.~\ref{n-1 expansion},~\ref{n-1 expansion 2},
\ba \notag
Z (n',k, t)& =   \\  \notag 1 + n'  &
\hspace{-3mm}
 \sum_{\substack{\mathcal{C}; \text{ one} \\ \text{marked loop} }}
 \hspace{-3mm}
 W_\mathcal{C} \,
(3 k)^\text{length}
(t/3)^\text{no. self-contacts} + \ldots.
\end{align}
Separating the sum into a sum over configurations of the marked loop (polymer) and a sum over the configurations of the other loops, and performing the latter,
\be
Z(n',k, t) =  1 + n'  \, Z_\text{pol} \lf k , t \ri + \ldots
\ee
We discuss only the partition function, but one may easily check that the natural geometrical correlation functions in the polymer problem, i.e. the watermelon correlators, can be expressed as correlators of $\vec S_\perp$ in the replica limit $n' \rightarrow 0$.

The polymer multicritical point at $(k,t) = (\f{1}{3},3)$ corresponds to the CPLC at $p=1/3$, $q=1/2$ and therefore to the sigma model with full $O(n)$ symmetry. Varying $(k,t)$ away from this point introduces symmetry-breaking perturbations, of which the most relevant are $\gamma_1$ and $\gamma_2$:
\ba\label{perturbed lagrangian for polymer}
& \mathcal{L}  = \f{K}{2} \lf (\nabla \vec S)^2 + \gamma_1 O_{\perp1} + \gamma_2 O_{\perp2} \ri, \\ \notag
O_{\perp1}  & = {\vec S_\perp}^2,  \quad\qquad O_{\perp2}  = (\vec S_\perp^2)^2 - \f{2(n+1)}{n+4} S_\perp^2.
\end{align}
Before considering the more detailed RG picture, we identify the phases in Fig.~\ref{polymer phase diagram} with the phases of this perturbed sigma model. These are characterized by whether the $\mathbb{Z}_2$ and $O(n')$ symmetries are broken, i.e. whether  $\langle S^1 \rangle \neq 0$ and whether  $\langle \vec S_\perp \rangle \neq 0$. 

When $\langle \vec S_\perp \rangle \neq 0$, the polymer fills the system densely and the transverse modes $\vec S_\perp$ are in a Goldstone phase. To see that such a phase is possible, note that when fluctuations in $S^1$ are massive we may imagine integrating them out to get  an effective $O(n')$ sigma model for $\vec S_\perp$. In the limit $n'\rightarrow0$, this sigma model has a Goldstone phase as a consequence of the beta function in Eq.~\ref{basic RG equation}. This is just the theory for the dense polymer with crossings studied in Ref.~\cite{dense loops and supersymmetry}.

The phase with $\langle S^1\rangle=\langle \vec S_\perp \rangle =0$ does not appear upon perturbing around the $\Theta$ point, since the latter is controlled by the fixed point at infinite stiffness. However the other three do:

{\bf Zero-density phase, $\langle \vec S_\perp \rangle = 0$, $\langle S^1 \rangle \neq 0$}. The leading effect of  reducing the length fugacity $k$ below the $\Theta$-point value is to introduce a mass for $\vec S_\perp$, so that $\vec S$  orders in the longitudinal direction and the transverse modes $\vec S_\perp$ are massive.  Correlators thus decay exponentially for the polymer, and it has a finite typical size.\footnote{In the zero density phase the \emph{un}marked loops remain in the Goldstone phase, being unaffected by the polymer since it is short. In order to write down watermelon correlation functions for the unmarked loops, we would have to extend $S^1$ to a vector $\vec S_1$ with $m\rightarrow 1$ components, increasing the symmetries of the field theory to $O(m) \times O(n')$.}

{\bf Dense phase with Ising disorder, $\langle \vec S_\perp \rangle \neq 0$ \& $\langle S^1\rangle=0$.} Moving away from the $\Theta$-point by increasing $k$ makes $S^1$ massive, and $\vec S$ `orders' in the transverse plane. The system thus goes from the  Goldstone phase of the $O(n\rightarrow 1)$ model to the Goldstone phase of the $O(n\rightarrow 0)$ model. The two-leg watermelon correlation function is a constant at long distances, as discussed in Sec.~\ref{correlation function section}, meaning that the polymer is dense.

{\bf Dense phase with Ising order, $\langle \vec S_\perp \rangle \neq 0$ \& $\langle S^1\rangle\neq 0$.} For appropriate values of the coefficients $\gamma_1$, $\gamma_2$,  the renormalised free energy is minimised when both $\langle \vec S_\perp \rangle$ and $\langle S^1\rangle$ are nonzero, so that both symmetries, $\mathbb{Z}_2$ and $O(n')$, are broken.

[A minor subtlety regarding the above classification is that only gauge-invariant operators are meaningful in the loop model/$\rp^{n-1}$ model. For this reason the global symmetry of the perturbed loop model\footnote{The zero-density phase breaks no symmetries of the perturbed $\rp^{n-1}$ model, the `dense phase with Ising order' fully breaks $O(n')$, and the `dense phase with Ising disorder' breaks only $SO(n')$.} is $O(n') = \mathbb{Z}_2 \times SO(n')$, rather than $\mathbb{Z}_2\times O(n')$ as in the perturbed $O(n)$ sigma model. However we are free to use the language of the latter, which (for the reasons discussed in Sec.~\ref{continuum description}) captures the universal properties of the perturbed $\rp^{n-1}$ sigma model in the regime we are considering.]

The field theory description also determines the nature of the phase transitions. First, consider the thin solid line in Fig.~\ref{polymer phase diagram}. The field $S^1$ is Ising-ordered on both sides of this transition, and its massive fluctuations play no role in the critical behaviour of $\vec S_\perp$. We therefore have the critical point of the $O(n')$  model in the limit $n'\rightarrow 0$. This is the usual description of the  self-avoiding walk \cite{replica for loops, cardy book}, confirming what we expect and find in the polymer problem.

Next, consider the dashed line in Fig.~\ref{polymer phase diagram}. Here, $S^1$ undergoes an ordering transition at which $\mathbb{Z}_2$ symmetry is broken. Thus we would expect an Ising transition. We must check, however, that the massless degrees of freedom associated with $\vec S_\perp$ --- which are in the Goldstone phase --- do not modify the Ising critical behaviour. But it is easy to see they do not. The most relevant coupling allowed by $\mathbb{Z}_2\times O(n')$ symmetry is via the product of the energy operators,
\be
E_{\text{Ising}} \times E_{\text{Goldstone}}.
\ee
This composite operator has dimension $(\text{length})^{-3}$, so is irrelevant.

Finally, consider the thick line in the figure, which separates phases breaking different symmetries ($\mathbb{Z}_2$  on one side, and $O(n')$ on the other). According to Landau theory this transition should be first order, as it is numerically found to be \cite{foster universality}.

What does Ising order/disorder mean for the polymer? Define a new configuration of Ising spins $\mu_F$ on the faces $F$ of the square lattice by the requirement that the polymer is the (only!) domain wall in this configuration. Then the dense phase with Ising \emph{disorder} corresponds to antiferromagnetic $\emph{order}$ in $\mu$, while $\mu$ is disordered in the dense phase with Ising order. (To show this, we write $\< \mu_F \mu_{F'}\>$ in terms a correlator of twist fields in the $\rp^{n-1}$ model, which force the fields $Q^{1a}$ with $a>1$ to change sign on a line connecting $F$ and $F'$.) Antiferromagnetic order in $\mu$ is equivalent to the `crystalline' order of Ref.~\cite{foster universality}, and it becomes perfect when the polymer visits every link of the lattice. It is also essentially equivalent to the Ising order  defined in Refs.~\cite{Blote Nienhuis, Nienhuis communication} for a different model.

For a more detailed picture of the phase diagram, we use the RG equations for the sigma model. We must include the lowest two anisotropies as in Eq.~\ref{perturbed lagrangian for polymer}, where $\gamma_1$, $\gamma_2$ are linearly related to the perturbations 
\ba
\delta k & = k-1/3,& \delta t &= t - 3
\end{align}
when these are small (we will give approximate expressions below). After running the RG up to a large time $\tau_*$, we have (for $n=1$)
\ba\notag
K_* & \sim \f{\tau_*}{2\pi}, &
\gamma_{1*} & \sim  \f{\gamma_1 e^{2\tau_*}}{(\tau_*/2\pi K)^2}, &
\gamma_{2*} & \sim  \f{\gamma_2 e^{2\tau_*}}{(\tau_*/2\pi K)^7}.
\end{align}
For generic small initial values $(\gamma_1, \gamma_2)$, we will renormalise to a regime where $\gamma_{1*} = O(1)$ and $|\gamma_{2*}| \ll |\gamma_{1*}|$,  putting us deep within one of the phases  -- either the zero density phase or the dense phase with Ising disorder, depending on the sign of $\gamma_1$. The phase transitions occur instead in the regime where the renormalised $\gamma_{1*}$ and $\gamma_{2*}$ become of order one simultaneously. 

Since the stiffness of the renormalised sigma model is large ($\tau_* \sim \ln |\gamma_1|^{-1/2} \sim  \ln |\gamma_2|^{-1/2}$) we may determine which phase it is in simply by minimising the potential in the renormalised Lagrangian. In doing this we must bear in mind the constraint $\vec S^2= 1$. We find the three phases described above, with the phase transition lines located at:
\ba\notag
\text{SAW}:&     & \gamma_{1}&\simeq \phantom{+}\f{4 }{5 } \gamma_{2} \lf \f{4\pi K}{\ln 1/|\gamma_2|} \ri^5 \quad & (\gamma_{2}&>0), \\ \notag
\text{Ising}:&    & \gamma_{1}&\simeq-\f{3}{5 }\gamma_{2} \lf \f{4\pi K}{\ln 1/|\gamma_2|} \ri^5  \quad & (\gamma_{2}&>0), \\\notag
\text{$1^\text{st}$ order}: &     & \gamma_{1}&\simeq-\f{1}{5}\gamma_{2} \lf \f{4\pi K}{\ln 1/|\gamma_2|} \ri^5 \quad &  (\gamma_{2}&<0). 
\end{align}
(These formulas are valid asymptotically close to the $\Theta$ point.) For a very crude estimate of the relation between $(\gamma_1, \gamma_2)$ and $(\delta k, \delta t)$ we can evaluate the right hand side of Eq.~\ref{polymer node} for spatially constant $\vec S_l$, and take the logarithm of the  Boltzmann weight for a node to obtain the potential terms in the bare Lagrangian. We find
\be
\gamma_1 \sim - C \left( \delta k + \f{2}{45}  \delta t \right), \qquad
\gamma_2 \sim - C \f{\delta t}{18},
\ee
where $C$ is an undetermined constant.

The above confirms that the three phases meet at the $\Theta$-point, and shows that the SAW and Ising critical lines are asymptotically parallel as they approach the $\Theta$ point. 

A remarkable consequence of the above field theory mapping is that the $\Theta$ point of the ISAT -- despite the simplicity and naturalness of this model -- is in fact an infinite order multicritical point! We have mentioned the two most relevant anisotropies $O_{\perp1}$ and $O_{\perp2}$, but there is an infinite number of these, with $O_{\perp k} = (S_\perp^2)^k + \ldots$, and generic perturbations of the ISAT $\Theta$ point will introduce all of them with couplings $\gamma_k$. At $n=1$, the RG equation for $\gamma_k$ is 
\be
\f{\dd \gamma_k}{\dd \tau} = \lf 2 - \f{2 k^2 -k + 1}{2\pi K} \ri \gamma_k + \ldots
\ee
so they are all relevant at the $K=\infty$ fixed point. 

The $\Theta$ point of the ISAT does not therefore represent the universality class of the \emph{generic} $\Theta$ point polymer model with crossings. (This explains why the model in Ref.~\cite{bedini trails with additional interactions} shows different behaviour to the ISAT.)  A similar argument explains why the three-dimensional ISAT \cite{Prellberg and Owczarek 3D} shows distinct universal behaviour from standard models of polymer collapse.

In two dimensions the generic $\Theta$-point behaviour in the presence of crossings is different from that \cite{duplantier saleur polymer} in their absence. Since the ISAT $\Theta$ point does not represent the generic behaviour of the $\Theta$-point polymer with crossings (which we expect to be described by the tricritical $O(N\rightarrow0)$ model), exact exponents for the latter are still unknown. We will discuss RG flows for $\Theta$-point polymers in detail in a separate publication.

\section{Outlook}
\label{conclusions}

The loop models we have discussed are described by `replica' sigma models of the kind familiar from localization and polymer physics. Such problems remain at the frontier of our understanding of critical phenomena, and we hope that the transitions in the loop models will provide a testing ground for new approaches.

While exact results for the critical behaviour discussed in Sec.~\ref{critical line} would be desirable, the development of more accurate analytical approximations would also be enlightening. On the numerical side,\footnote{The IPLC is also a candidate for numerical simulations: we expect universal behaviour to coincide with that of the CPLC, but this remains to be confirmed.} work on the critical loop model should  be extended to other values of $n$, either via Monte Carlo or the transfer matrix \cite{Ikhlef nonintersection}, in order to pin down the properties of the whole family of critical points for $0<n<2$. We plan to return to  these issues. (Three-dimensional $\rp^{n-1}$ loop models exist as well --- we will report numerical results elsewhere.)

The connection between the CPLC at $n=1$ and disordered fermions remains an open question. To begin with, recall the situation for the loop model without crossings ($p=0$). This can be related to localisation in at least two ways. Firstly, a limiting case of the Chalker Coddington model for the quantum Hall effect \cite{chalker coddington}, in which the scattering matrices at a node become `classical', yields the  loop model without crossings --- i.e. classical percolation. This is the familiar semiclassical description of the quantum Hall transition \cite{trugman}, but because quantum tunnelling has not been taken into account, it does not correctly capture the universal critical behaviour. However, the loop model has a second relationship with localisation which is less obvious and which does not rely on suppressing quantum tunnelling. This is due to an exact mapping from a network model for the spin quantum Hall transition (an analogue of the quantum Hall transition, but in symmetry class $C$ rather than $A$) to the loop model \cite{glr, beamond cardy chalker, Mirlin Evers Mildenberger, cardy network models review}.

For loops with crossings ($p>0$) we can again construct a mapping of the first kind by taking a `classical' limit in a network model with a Kramers doublet on each edge (replacing each quantum node with one of the three classical possibilities in Fig.~\ref{decompnode}). However this correspondence is rather trivial because of the explicit suppression of quantum tunnelling. It would be interesting to know whether the analogy with localisation goes beyond this --- in particular, whether the critical behaviour of the loop model can be related to the critical behaviour of a true localisation problem. (It is interesting to note that our value of $\nu$ is close to estimates of $\nu$ for the symplectic class \cite{symplectic class numerics}.)

Returning to loop models in their own right, there is a good understanding of the zoology of critical points in loop models without crossings, many of which fit into the one-parameter family of universality classes in SLE$_\kappa$. In general, crossings take us outside this family. Here we have discussed a new line of critical points exemplifying this, but we certainly do not expect that this exhausts the possibilities for new critical behaviour --- much remains to be learned.

\section*{Acknowledgements}

We are very grateful to J. Chalker for many useful discussions, and for valuable comments on the manuscript. AN also thanks L. Fu, J. Jacobsen, L. Jaubert, A. Ludwig, T. Prellberg, K. Shtengel and especially J. Cardy and F. Essler for useful discussions, and B. Nienhuis for correspondence.  PS, AMS and MO acknowledge financial support from the Spanish DGI and FEDER, grant No. FIS2012-38206 and No. AP2009-0668. This work was supported in part by the EPSRC.

\end{document}